\documentclass[usegraphicx,usenatbib]{mn2e}

\begin{document}

\title[Iron line reverberation in NGC 4151]{Modelling the broad Fe K$\alpha$ reverberation in the AGN NGC~4151}

\author[Cackett et al.]{E.~M.~Cackett$^1$\thanks{ecackett@wayne.edu}, 
A.~Zoghbi$^{2, 3}$,
C.~Reynolds$^{2, 3}$,
A.~C.~Fabian$^4$, 
E.~Kara$^4$,
P.~Uttley$^5$,
\newauthor
and
D.~R.~Wilkins$^6$
\\$^1$ Department of Physics \& Astronomy, Wayne State University, 666 W. Hancock St., Detroit, MI 48201, USA
\\$^2$ Department of Astronomy, University of Maryland, College Park, MD
20742, USA
\\$^3$ Joint Space-Science Institute (JSI), College Park, MD 20742-2421, USA
\\$^4$ Institute of Astronomy, University of Cambridge, Madingley Rd,
Cambridge, CB3 0HA
\\$^5$ Astronomical Institute `Anton Pannekoek', University of Amsterdam,
Science Park 904, 1098 XH Amsterdam, the Netherlands
\\$^6$ Department of Astronomy \& Physics, Saint MaryÕs University, Halifax, NS, B3H 3C3, Canada}

\date{Received ; in original form }
\maketitle

\begin{abstract}
The recent detection of X-ray reverberation lags, especially in the Fe~K$\alpha$ line region, around Active Galactic Nuclei (AGN) has opened up the possibility of studying the time-resolved response (reflection) of hard X-rays from the accretion disk around supermassive black holes.  Here, we use general relativistic transfer functions for reflection of X-rays from a point source located at some height above the black hole to study the time lags expected as a function of frequency and energy in the Fe K$\alpha$ line region.  We explore the models and the dependence of the lags on key parameters such as the height of the X-ray source, accretion disk inclination, black hole spin and black hole mass.  We then compare these models with the observed frequency and energy dependence of the Fe K$\alpha$ line lag in NGC~4151.  Assuming the optical reverberation mapping mass of $4.6\times10^7~M_\odot$ we get a best fit to the lag profile across the Fe K$\alpha$ line in the frequency range $(1-2)\times10^{-5}$ Hz for an X-ray source located at a height $h = 7^{+2.9}_{-2.6}~R_G$  with a maximally spinning black hole and an inclination $i < 30^\circ$. 

\end{abstract}

\section{Introduction}

Over the last 30 years reverberation mapping \citep{blandmckee82} at optical wavelengths has become firmly established as one of the premier ways to measure the mass of supermassive black holes in Active Galactic Nuclei (AGN) \citep[e.g.][]{petersonetal04}.  It uses a simple principle - a primary variable signal originating from close to the black hole irradiates gas further out leading to a reprocessed variable signal that is a delayed, and smeared out version of the primary signal.  In other words, there are two correlated light curves, with the reprocessed light curve delayed with respect to the direct light curve.  The time delay, or `lag',  is caused by the difference in light travel time between the primary emission which travels directly to the observer, and the reprocessed emission which travels an extra distance.  The lag, then, leads to a measurement of the physical size of the emission region.   In optical reverberation mapping, the reprocessed signal takes the form of broad emission lines, with the time lags indicating that the broad emission line region is typically tens of light-days in size.  Of course, the dynamics of the gas in the broad emission line region sets the shape of the line profile, thus combining the size of the emitting region from the measured lag, and the velocity of the broad line gas from the width of the emission lines leads to an estimate of the black hole mass.  In this way, masses for more than 50 supermassive black holes in AGN have been successfully measured \citep{petersonetal04,bentz09}.  Furthermore, scaling relations derived from optical reverberation mapping results allow for black hole mass estimates from single-epoch spectra over a wide range in redshift \citep[e.g.,][]{kaspi00,vestergaard02,bentz09_rl}.  Moreover, the best available data are now allowing the structure of the broad line region to be probed \citep{bentz10}.

In a similar fashion, reflection occurring in the X-ray band should also lead to reverberation.  While optical reverberation mapping probes the AGN on size scales of light weeks to light months, the X-ray emitting region is much closer to the black hole, probing within a few hundred light seconds.  The expectation of X-ray reverberation comes from one of the standard models used to explain the X-ray spectra of AGN \citep[e.g.][]{georgefabian91}.    In this reflection model, hard X-rays from the corona (the power-law component), irradiate the accretion disk, leading to a reflected component consisting of fluorescent emission lines and scattered continuum emission. For detailed reviews of X-ray reflection, see, for example, \citet{reynoldsnowak03,miller07}.  As this reflected emission originates in the accretion disk, the dynamical and relativistic effects present there are imprinted on the emission.  The strong relativistic Doppler effects skew and broaden the emission lines, while the gravitational redshift further broadens and shifts it to lower energies.  The result is an emission line with a characteristic broad and skewed profile, with the exact shape sensitive to how far the inner accretion disk extends into the gravitational potential well of the black hole \citep{fabian89}.  

The most prominent of the fluorescent emission lines (due to abundance and fluorescence yield) is the well-known Fe K$\alpha$ line, which occurs at 6.4 keV (or higher, depending on ionization state).  This line is widely observed in AGN \citep[e.g.,][]{reynolds97,nandra97, nandra07}, and often displays the characteristic broad, skewed profile expected for emission from within a few gravitational radii of the black hole \citep[e.g.,][]{tanaka95}.  As the location of the innermost stable circular orbit (ISCO) around the black hole is set by the black hole spin (from 1.23 $R_g$ for a maximally spinning black hole to 6 $R_g$ for a non-rotating black hole), the measure of the inner accretion disk radius from spectroscopic modeling of the broad Fe~K$\alpha$ line provides a route to measuring the black hole spin in AGN and stellar-mass black hole X-ray binaries \citep[e.g.][]{brenneman06,miller09}.

Another prominent feature of the X-ray spectra of many AGN is the so-called `soft-excess'.  This broad feature occurs at soft X-ray energies, and is typically well-characterized by a blackbody with a temperature of approximately  $kT = 0.1 - 0.2$ keV.  The nature of the soft excess has been the subject of significant debate \citep[e.g.,][]{crummy06, gierlinski04}, though, the fact that the shape of the soft excess is independent of black hole mass and luminosity, points to an atomic origin for the emission.  It is naturally explained in the reflection paradigm -- many emission lines at soft energies should be produced by reflection, and the same relativistic effects that broaden and skew the Fe K~$\alpha$ line will also be present, blurring the many lines into a broad, blackbody-like emission component.

The reflection model, therefore, has a number of key predictions regarding the reverberation signal one would expect to see.  There should exist a lag between the primary irradiating component and the reflection components. In other words, one would expect to see  emission from both the Fe~K$\alpha$ line and the soft excess lagging the power-law emission.   As reflection is thought to originate from just a few gravitational radii of the black hole, in this picture, X-ray reverberation lags should correspondingly be of the order of a few tens to hundreds of seconds ($1 R_G/c = 49$ seconds for $M = 10^7$~M$_\odot$). 

Within the last few years, the first handful of detections of X-ray reverberation have now been made.  The breakthrough in the search for X-ray reverberation came from studying the narrow-line Seyfert 1 AGN, 1H~0707$-$495 \citep{fabian09}.  1H~0707$-$495 is a bright, highly-variable AGN, which has a particularly high iron abundance.  This leads to its X-ray spectrum showing not just a strong Fe~K emission line, but also a strong Fe~L line.  Fe~L occurs at $\sim$0.7 keV, an energy close to where the effective area of typical X-ray telescopes peaks, leading to a significantly higher count rate than at 6.4 keV.  Calculating the lag between the region of the spectrum dominated by the power-law (1 -- 4 keV) and the Fe L line (0.3 -- 1.0 keV) revealed the Fe L line lagging behind the power-law component (referred to as a `soft lag') by approximately 30 seconds on short timescales, $(1 - 2)\times10^{-3}$ Hz.  However, on much longer timescales ($2\times10^{-4}$ Hz) a lag in the opposite sense  (a `hard lag') of order a few hundred seconds was found.  While the hard lags have been seen before in AGN and X-ray binaries \citep[e.g.][]{papadakis01, miyamoto89}, this was the first ever significant detection of a soft lag in an AGN, as expected from reverberation from the inner accretion disk.  The magnitude of the lag, just a few tens of seconds, indicates the emitting region is just a few tens of light-seconds from the primary X-ray source, which corresponds to less than a few $R_G$.

Thus, 1H~0707$-$495 displays frequency-dependent lags between the power-law and Fe L bands, and it shows a transition between two types of lags -- hard lags on long timescales and soft lags on shorter timescales.  Hard lags are well-known in black hole X-ray binaries \citep[e.g.][]{nowak99, kotov01}, and are thought to arise due to viscous propagation of mass accretion fluctuations in the disk \citep{uttley11}.  The soft lags, on the other hand, can easily be explained by X-ray reflection from the inner accretion disk, and this model also predicts how one would expect the lags to evolve across the energy spectrum.  Instead of simply looking at the frequency-dependence of the lag between two bands, one can also look at the energy-dependence of the lags.  Thus, in a given frequency range, one can calculate the lag in every energy-band across the spectrum with respect to a lightcurve in a reference band.  This then displays the energy-dependence of the lag across the spectrum, showing the relative lag between the energy bands.  This provides extra vital information -- in the reflection model at the frequencies where the soft lag is apparent,  one would expect the regions of the spectrum dominated by the reflection component to show the largest lag compared to the region where the power-law dominates.  So, naively, one would expect the lags to roughly follow the reflection fraction \citep[reflection flux / power-flux, see][for more details]{zoghbi13}.  This is exactly what is seen in 1H~0707$-$495 \citep{zoghbi10,zog11_1h0707,kara13a}.  

Following on from the detection of soft lags in 1H~0707$-$495, it became apparent that soft lags should be searched for between the power-law dominated region of the spectrum and the soft excess in other highly variable AGN.  These searches were fruitful, and soft lags were detected in many more objects \citep{zog11_rej1034, emma11, tripathi11, demarco11, demarco13,cackett13,fabian13, alston13}.  Interestingly, now that there is a reasonable sample size, one can look at how the lag evolves with the black hole mass.  As the size-scale of the black hole is set by its gravitational radius, and hence its mass, we should expect the magnitude of the soft lag to increase linearly with black hole mass.  Another important consideration is the frequency at which the soft lag occurs.  This, too, is set by the size scale of the region, and hence the black hole mass, so, we would expect the frequency of the soft lag to decrease linearly as the black hole mass increases.  A systematic analysis by \citet{demarco13} detected soft lags in 15 AGN and shows that both these mass scalings are seen, as expected.  Just as optical reverberation mapping scaling relations have led to single-epoch mass estimates, this opens the possibility of single-epoch mass estimates based on X-ray soft lag measurement.

Although many soft lags are now seen, a detection of broad Fe K reverberation was still lacking.  This changed with the recent analysis of NGC 4151 \citep{zoghbi12}.  NGC 4151 is the brightest AGN through the Fe K band, 300 times brighter than 1H~0707$-$495 at 6.4 keV. It is therefore the ideal object to search for broad Fe K reverberation.  In \citet{zoghbi12} we searched for lags across the energy band, finding lags of order 3000s on timescales of $10^{5}$s between the 5--6 keV band and the 2--3 keV and 7--8 keV bands.  The mass of NGC 4151 is estimated as $M = (4.6 \pm 0.5)\times10^7$  M$_\odot$ \citep{bentz06}, so this lag corresponds to approximately $13~R_G$.  Interestingly, the broad lag profile resembles the shape of a relativistically-broadened iron line.  Moreover, we also find that the lag profile evolves with frequency -- the peak of the profile shifts to lower energies at higher frequencies, consistent with the red wing of the line being emitted at smaller radii (and hence seen at higher frequencies), as expected from reflection off the inner disk.   Since the discovery of broad Fe K lags in NGC~4151 we have also seen Fe K lags in 6 more objects: MCG-5-23-16 and NGC 7314 \citep{zoghbi13}, 1H~0707$-$495 \citep{kara13a}, IRAS~13224$-$3809 \citep{kara13b}, Mrk 335 and Ark 564 \citep{kara13c}.  The Fe K lags in these 7 objects also appear to scale with mass \citep{kara13c}.  

While the lags show all of the properties expected for reflection from the inner accretion disk, it is important to assess the uniqueness of this interpretation.  In fact, an alternative model for the hard and soft lags has been proposed \citep{lancemiller10,legg12}.  In this competing model, both the hard and soft lags are caused by scattering of X-rays passing through an absorbing medium that partially covers the source.  In this model the soft lags are just an artifact of a transfer function with a sharp edge causing oscillatory lags.  The absorbing material is located much further away from the black hole (hundreds to thousands of $R_G$) compared to the relativistic reflection scenario.  This model is able to successfully reproduce the frequency-dependent lags \citep[][ though see \citealt{emma11} who find a statistically better fit from the relativistic reflection scenario]{lancemiller10}.  Detailed arguments against this interpretation are presented in \citet{zog11_1h0707}. Furthermore, as yet, the energy dependence of this model has not been explored.  However, it was noted by \citet{kara13c} that in both Ark 564 and Mrk 335 the energy dependence of the lags is very different at high and low frequencies.  While at high frequencies Fe K reverberation is seen, at low frequencies the lag-energy spectrum is smooth showing no signs of reverberation.  On the contrary, if distant reflection is the origin of the lags, then there should be a clear signature of reflection on the longest timescales, which is not seen.  The distant reflection interpretation is not the focus of this work, and we will not discuss it further.

Thus far, much of the focus on these Fe~K lags has been in obtaining robust detections.  However, the next step is to start to explore the implications for the geometry and kinematics of the inner region around the black hole based on these lags, in the same way that optical reverberation mapping has progressed from lag measurements to more detailed studies.  Here, we take the first step toward a more detailed understanding of the specific frequency and energy dependence of Fe K lags by comparing the lags seen in NGC~4151 with general relativistic transfer functions calculated for reverberation from the inner accretion disk.  In Section~\ref{sec:grtf} we describe the model transfer function.  In Section~\ref{sec:freqevo} we explore the frequency dependence of the lags from these models and compare it to NGC~4151.  In Section~\ref{sec:enevo} we explore the energy dependence of the lags from these models and again compare it to NGC~4151.  Finally, we present our conclusions in Section~\ref{sec:discuss}.

\section{General relativistic transfer function}\label{sec:grtf}

\begin{figure*}
\centering
\includegraphics[width=12cm]{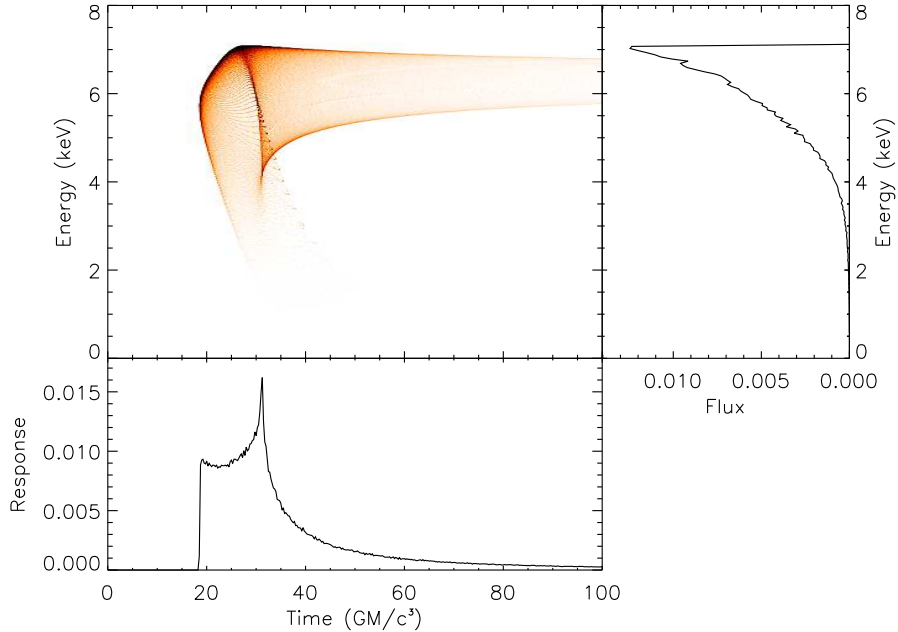}
\caption{{\it Main panel}: 2-dimensional transfer function for a disk around a maximally spinning black hole with $i = 45^\circ$, and $h = 10$ GM/c$^2$. {\it Right panel:} The time-average response of the same disk.  This is the familiar relativistic emission line profile.  {\it Bottom panel:} The energy-averaged response, showing the response of the entire emission line as a function of time.}
\label{fig:tf}
\end{figure*}

The time-averaged response of an accretion disk around a black hole is well-studied, and is just the reflection spectrum \citep[e.g.,][]{rossfabian05}.  As we describe above, the energy spectrum caused by reflection from the inner accretion disk is set by the kinematics and relativistic effects occurring in the emission region, as well as the ionization state of the accretion disk, and the geometry of the hard X-ray source and disk.   Reverberation, however, looks at how the reflected emission responds to changes in the driving continuum emission.  Thus, it is the time-resolved response of the disk that matters.  As in the time-averaged case the kinematics, relativistic effects and ionization state all contribute to the observed energy spectrum.  However, the time at which each part of the accretion disk responds to some change in the hard X-ray emission is primarily set by the distance between these two regions as well as additional effects such as the apparent delay of the travel of light through a region of curved spacetime close to the black hole according to the time measured by an observer at infinity \citep[`Shapiro delay',][]{shapiro64}.  This information is all encoded in the transfer function which has been well explored in the past \citep{campana95,reynolds99,kotov01,poutanen02,nayakshin02,cassatella11,wilkins13}.  

Here, we use the general relativistic transfer functions as calculated by \citet{reynolds99}.  The model assumes a simple lamppost geometry, with a point-like source of irradiating hard X-ray photons located at a height, $h$, above the black hole.  The accretion disk is assumed to be prograde, and here we consider two cases for the black hole spin, a slowly rotating black hole ($a=0.1$) and a maximally-spinning Kerr black hole ($a=0.998$).  Other parameters in the model include the inclination, $i$, and the outer accretion disk radius, $R_{\rm out}$.   The calculations are described in detail in \citet{reynolds99} and we refer the interested reader there for an in-depth discussion \citep[see also][for a discussion of GR transfer functions relating to reverberation]{wilkins13}.  The only change from \cite{reynolds99} is the treatment of the outer disk radius.  We compute the strong-field GR transfer functions out to $R_{\rm out} = 100$ GM/c$^2$, and this is the outer radius used in \citet{reynolds99}.    However, here we extend the outer radius to $R_{\rm out} = 1000$ GM/c$^2$ by using a weak-field analytical approximation to compute the transfer function for $100 < R  \le 1000$ GM/c$^2$.  This significantly speeds up the computation time and is a very good approximation given that the GR effects are strongest on size-scales much less than $100$ GM/c$^2$.  All transfer functions used here are calculated in this way, with $R_{\rm out} = 1000$ GM/c$^2$.

In Figure~\ref{fig:tf} we show an example two-dimensional transfer function \citep[see][for more examples]{reynolds99} for a maximally spinning Kerr black hole with $h = 10$ GM/c$^2$, $i = 45^\circ$, and black hole mass,~$M$.  Summing the 2D transfer function along the time axis gives the familiar time-averaged Fe~K$\alpha$ line profile \citep{fabian89,laor91}, whereas summing the 2D transfer function over all energies gives the energy-averaged emission line response to a delta-function flare above the accretion disk.

\begin{figure}
\centering
\includegraphics[angle=270, width=8cm]{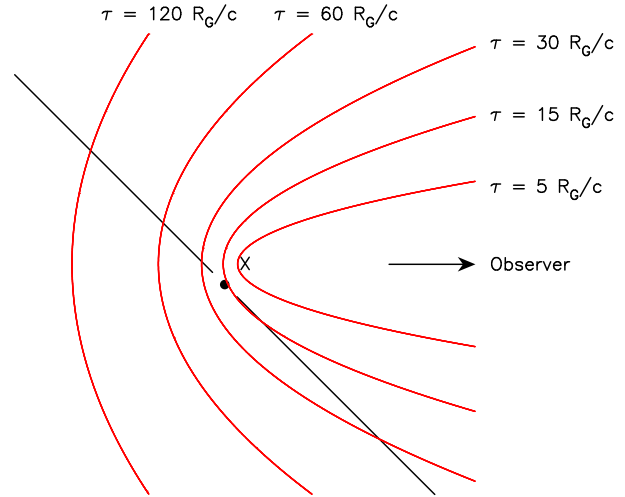}
\caption{Isodelay surfaces (red lines) at a time $\tau$ after a delta-function flare from an X-ray source (X) at height $h = 10~R_G$ above a non-spinning black hole (black circle), and an accretion disk (black line) inclined at 45$^\circ$ to the observer.  This figure does not include relativistic effects, and is based only on path-length considerations.}
\label{fig:isodelay}
\end{figure}

We briefly describe the shape of the 2D transfer function \citep[see also][for similar discussions]{reynolds99,wilkins13}.  For the Fe~K$\alpha$ emission line at 6.4 keV, the energy at which reprocessed photons are emitted is set by the dynamical and relativistic effects in the region of the disk where the reprocessing takes place \citep{fabian89,laor91}.  The times when reprocessed emission is seen is set by both the path length difference between direct emission from the source of irradiating photons and region on the accretion disk where reprocessing is taking place, as well as Shapiro delays set by photons traveling through curved space-time.  In Figure~\ref{fig:isodelay} we show the isodelay surfaces for a X-ray source at a height $h = 10~R_G$ above a black hole, with the accretion disk at an angle of $45^\circ$ to the observer.  At a time $\tau$ after a delta-function flare, we see emission from along a given isodelay surface.  Where the isodelay surface interacts with the disk indicates the radii of the disk we see at time $\tau$ (note this figure does not take relativistic effects into account).

Unless the accretion disk is face-on ($i = 0^\circ$), an observer will see reprocessed emission due to a flare from a central X-ray source above the disk from the near-side of the accretion disk first, due to the shorter path length to the observer (as can be seen from the isodelay surfaces in Figure~\ref{fig:isodelay}). Therefore the initial rise in the transfer function (shown in Figure~\ref{fig:tf}) occurs due to emission from the near-side of the accretion disk with the shortest path lengths.  
 The time and energy at which this occurs depends on the radius in the disk at which the reflected photons were emitted.  The lowest energies seen in an emission line (the `red wing') are emitted from the innermost part of the accretion disk where the effects of gravitational redshift and Shapiro delay are the strongest. Therefore, the initial rise comes at increasingly later times for the lowest energies partly due to increasing Shapiro delay, and partly due to path length effects.   The second peak in the transfer function occurs when light arrives from the far-side of the accretion disk.   As time after the flare increases, emission from outer radii only are seen, and hence a much narrower transfer function is seen at late times -- the envelope is set by Doppler broadening at the corresponding radii.

The 2D transfer function is sensitive to a number of parameters.  For instance, the inclination, height of the X-ray source above the black hole and the inner edge of the accretion disk (set by the black hole spin).  The behaviour of the transfer function as these parameters change is discussed in \citet{reynolds99}.  We briefly describe the effect of each parameter here.  Increasing the viewing inclination has the effect of broadening the transfer function.  For a face-on disk the transfer function is narrow as the path length to a given disk radius is the same for all azimuthal angles.  As the inclination increases one side of the disk is closer than the other, leading to a wide range of path lengths.  Furthermore, the line of sight velocities increase as inclination increases leading to larger Doppler broadening for higher inclinations.  

The height of the X-ray source above the disk also significantly changes the transfer function.  A smaller height leads to shorter overall time lags due to the path length decreasing.  Moreover, for smaller heights the general relativistic effects are significantly stronger.  This gives both greater gravitational redshifts and also larger time dilation.  Black hole spin also changes the 2D transfer function since spin changes the location of the innermost stable circular orbit (ISCO), with a maximally spinning black hole ($a = 0.998$) having an ISCO at GM/c$^2$ compared to 6 GM/c$^2$ for a non-spinning ($a = 0$) black hole \citep{bardeen72,thorne74}.  The presence of a disk closer to the black hole for higher spin leads to larger gravitational redshifts, and more prominent reprocessing at lower energies.  For examples of how inclination and black hole spin change the 2D transfer function, see figures in \citet{reynolds99}.

Finally, both the size and time scales are proportional to the black hole mass -- size scale goes like $R_G = GM/c^2$, while the light travel times go like $R_G/c = GM/c^3$ .  Therefore, the transfer function can be trivially scaled for any black hole mass.  In the following sections we discuss the frequency dependence of the lags and energy dependence of the lags based on these transfer functions, and compare to the observations of lags in NGC~4151.

\section{Frequency dependence of the lags}\label{sec:freqevo}

Here we consider how the time-evolution of the transfer functions corresponds to frequency dependence of the lags.  In this section we first consider the frequency dependence of lags from the entire Fe K emission line and discuss how inclination, height of the X-ray source, spin and black hole mass set the lags.  We then compare the observed frequency dependence in NGC~4151 with these models.  A similar discussion of such parameters on the frequency-dependence of the lags can be found in \citet{wilkins13}.  We discuss again here in order to clearly compare with observations of NGC~4151. 

To calculate the frequency-dependence of the lag we take the Fourier transform of the transfer function, assuming some reflected response fraction due to contribution from both the irradiating source and reflection component in the spectrum.  If we call $\psi(\tau)$ the transfer function in the time domain,  and define the reflected response fraction, $R$, as the (reflection flux)/(power-law flux), then, the transfer function (scaled by the reflected response fraction) in the frequency domain is $\Psi(f) = R \int_0^\infty \psi(\tau) e^{-i 2\pi f\tau} d\tau$, i.e. the Fourier transform of $\psi(\tau)$ scaled by $R$.  We calculate the phase difference, $\phi$, using the following equation:

\begin{equation}
\phi(f) = \tan^{-1}\left(\frac{\Im(\Psi)}{1 + \Re(\Psi)}\right)
\label{eq:phase}
\end{equation}
where $ \Re(\Psi)$ and $\Im(\Psi)$ are the real and imaginary parts of $\Psi$.  The time lag is then calculated from $\phi/2\pi f$. The $1$ on the denominator of Equation~\ref{eq:phase} is due to the contribution of the driving lightcurve in the reflected energy band.  The lightcurve being in both bands will add a zero lag contribution, i.e. will not change the imaginary part, but will act to dilute the lag and hence a change in the real part.  Under our definition, $R = 1$ would give equal contribution of the power-law and reflected fluxes in the energy band where we are measuring the lag.  Our definition of $R$ assumes no contribution from the reflected lightcurve in the power-law band, though note doing so would just further dilute the lag measured and would be equivalent to reducing $R$ \citep[see][]{kara13a,wilkins13}.

 For all but one of the following examples we assume $R$ across the Fe~K$\alpha$ line of 1, corresponding to equal contributions from both the power-law and Fe~K$\alpha$ line.  We achieve this by normalizing the area under the entire transfer function to 1, and hence $R$ corresponds to the reflected response fraction averaged over the Fe~K$\alpha$ line.  In the last example we discuss how changing the reflected response fraction changes the the absolute value of the size of the lag.

The frequency dependence of the lags shown in the following sections all have the same basic shape.  At low frequencies they show a positive lag between the Fe~K$\alpha$ line (reflected emission) and the primary emission.  This positive lag indicates that the reflected emission lags behind the primary emission, as expected.  Phase wrapping at high frequencies (discussed in more detail below) causes the lags to go to, and oscillate around, zero.  We stress that it is the positive lag that we identify as the reverberation signal detected in NGC~4151, and that the resemblance to the lags seen between the power-law and soft excess in other AGN, such as 1H~0707$-$495, is coincidental (in that source and with those energy ranges it is the negative lag that represents the reverberation signal).  Finally, note that we assume there are no continuum lags (often referred to as `hard lags') in NGC~4151.  This is supported by the zero lag between the 2 -- 3 keV and $>7$ keV bands at the frequencies where we see Fe~K$\alpha$ reverberation.

\subsection{Height of the X-ray source}
\begin{figure*}
\centering
\includegraphics[width=16cm]{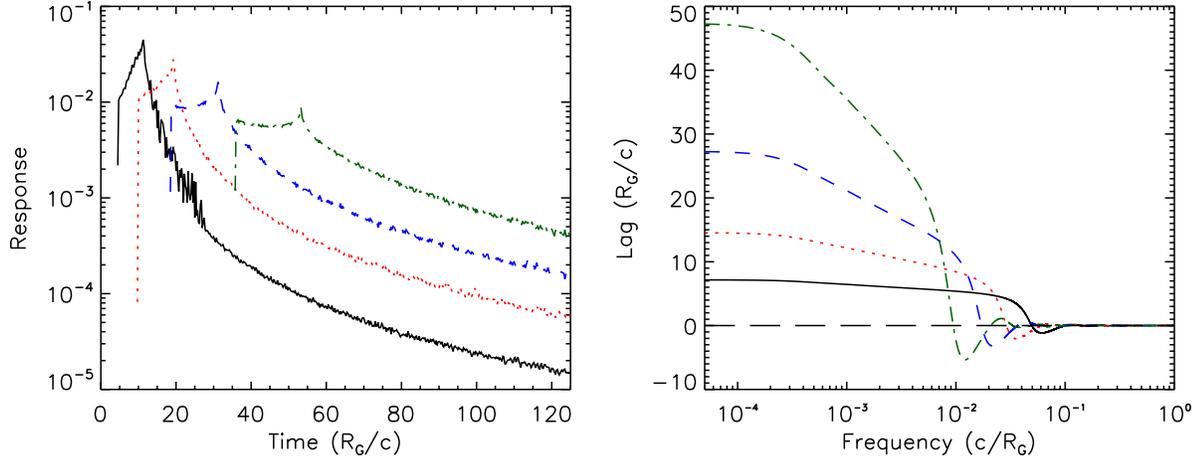}
\caption{{\it Left:} Transfer functions for an X-ray source located at height, $h$ above the disk.  The lines correspond to $h = 2 R_G$ (solid, black), $5 R_G$ (dotted, red), $10 R_G$ (dashed, blue), $20 R_G$ (dot-dashed, green).  {\it Right:} Lag vs frequency corresponding to the transfer functions shown in the left panel.  A positive lag is the Fe K line lagging behind the driving lightcurve, i.e. reverberation.}
\label{fig:h}
\end{figure*}

First, we discuss the height of the irradiating X-ray source above the black hole.  We assume a simple lamppost model, with an X-ray point source some height, $h$, above the black hole.  In Figure~\ref{fig:h} we show the transfer function along with the corresponding lags for four different heights above the disk ($h = 2, 5, 10$ and $20~R_G$).  We assume a maximally spinning black hole and an inclination of 45$^\circ$ in each case.  Two changes are noticeable for different heights --  firstly, the lag at the lowest frequencies is larger for greater heights and secondly, the frequency where phase wrapping occurs (the lags roll over and start to oscillate) decreases with increasing height. These are both caused by the fact that an increase in height leads to an increase in the overall time lags (as can be seen in the transfer functions).  

It is also worth discussing the general shape of the lag versus frequency plots.  The maximum lag reached is set by the mean of the transfer function (the average lag from the whole disk) and the reflected response fraction (discussed later).  Note that at low frequencies the lags flatten off and become constant with decreasing frequency.   At frequencies corresponding to periods less than twice the reverberation delay (set by the size of the region) the phase wraps around.  Since the phase is defined to be in the range $-\pi$ to $\pi$ one is unable to distinguish whether the phase shift is half a wave forwards or backwards i.e. a lag of $-\pi/2$ or a lead of $3\pi/2$.  In all our transfer functions we use an outer disk with a radius of 1000 $R_G$, thus phase wrapping from this region starts to occur at frequencies higher than about $1/2000 = 5\times10^{-4} c/R_G$ in the face-on case, and $1/4000 = 2.5\times10^{-4} c/R_G$ for completely edge-on.  This is the effect causing the decrease of the average lag starting a $2\times10^{-4}$ Hz.  One can also see in Figure~\ref{fig:h} that there is also a significant change in the slope (a second break) in the lags at a higher frequency.  If we examine the transfer function for $h = 10~R_G$, this second break corresponds to the significant drop in the transfer function at $t \sim~35 R_G/c$, and hence a change in the lags at a frequency of about $1/70 = 1.4\times10^{-2} c/R_G$ (for the model with $h = 10~R_G$).   Most of the response comes from the disk between about $t = 20 - 35 R_G/c$, and so this is where we see a rapid change in the measured lag.  At longer times (lower frequencies), the response comes from further out in the disk, where the emissivity is dropping off and so the average lag from the disk up to a given radius increases more slowly than on shorter times where the emissivity is highest and the average lag will change significantly with increasing radius.

\subsection{Inclination}
\begin{figure*}
\centering
\includegraphics[width=16cm]{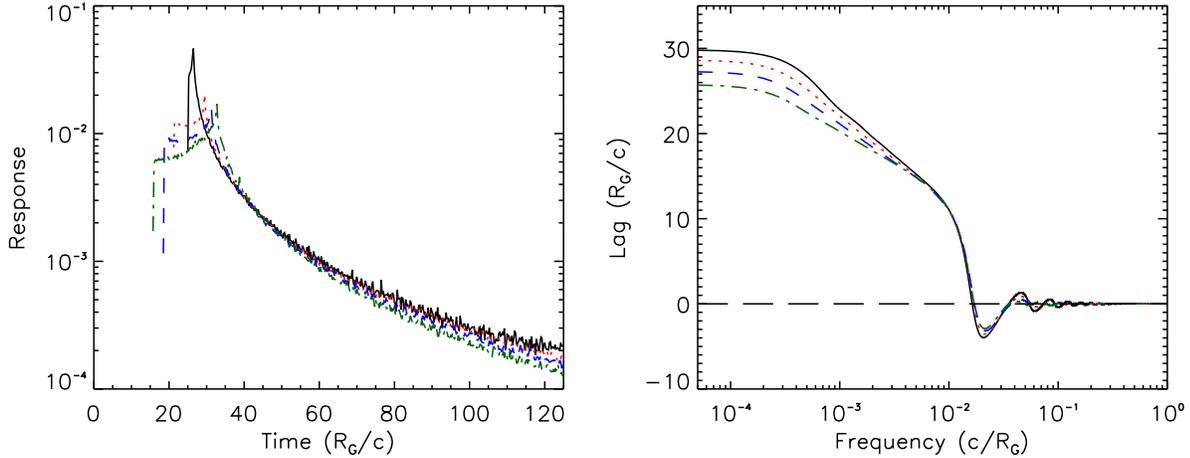}
\caption{{\it Left:} Transfer functions for an X-ray source located at a height of 10 $R_G$ above the black hole, but with different inclination angles.  The lines correspond to inclinations of $i = 5^\circ$ (solid, black) $30^\circ$ (dotted, red), $45^\circ$ (dashed, blue) and $60^\circ$ (dot-dashed, green).   {\it Right:} Lag vs frequency corresponding to the transfer functions shown in the left panel.}
\label{fig:inc}
\end{figure*}

Next, we consider the effects of source inclination.  Figure~\ref{fig:inc} shows the transfer functions and corresponding lags for inclinations of $i = 5, 30, 45$ and $60^\circ$.  All the models have $h = 10 R_G$ and assume a maximally spinning black hole.  The transfer function is narrowest for inclinations close to face-on, and increases in width with increasing inclination as the difference in path-length between photons being reprocessed on the near and far-side of the disk increases.  As can been seen in the right-hand panel of Figure~\ref{fig:inc}, this only has a small effect on the maximum lag, and does not have a noticeable effect on the frequency where phase wrapping occurs.  Note, however, that the maximum lag is longest for the lowest inclination, because the mean lag from the disk is longest as neither side of the disk is closer to the observer.  There is a more significant difference when looking at the effect of inclination on the energy dependence of the lags, which we will discuss later in section~\ref{sec:enevo}.

\subsection{Black hole spin}
\begin{figure*}
\centering
\includegraphics[width=16cm]{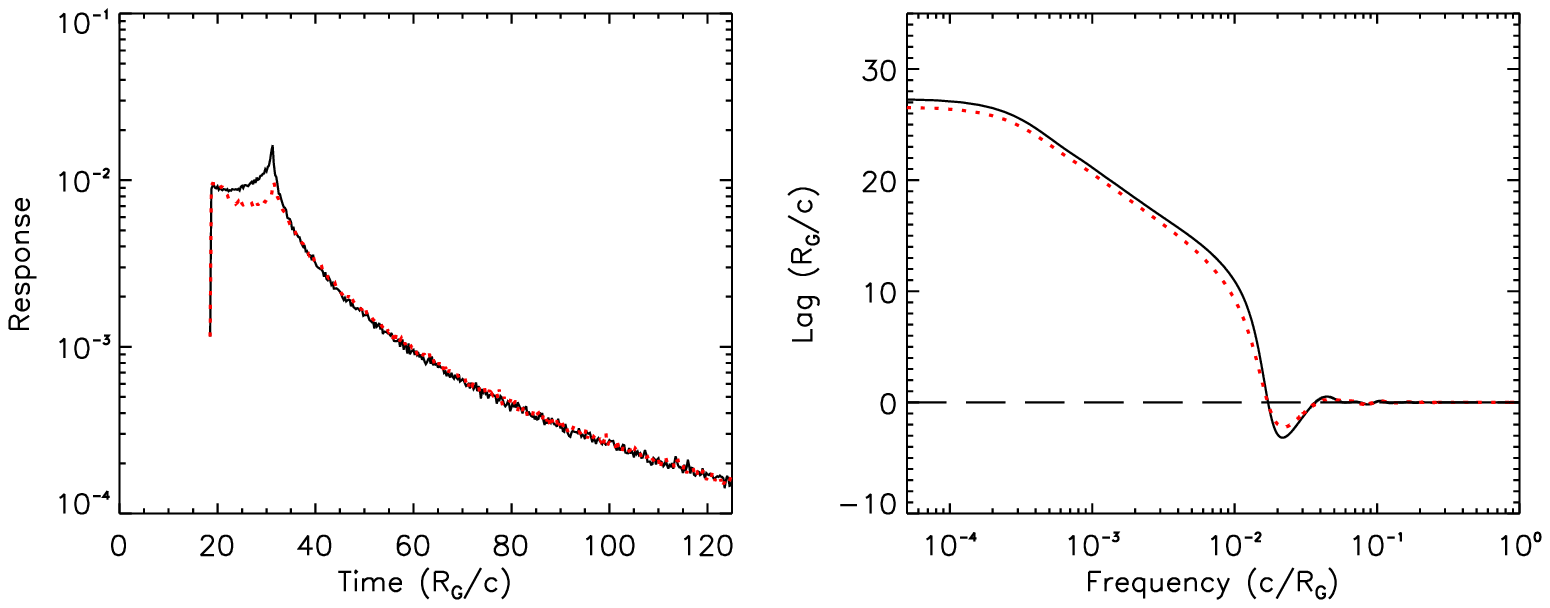}
\caption{{\it Left:} transfer functions for an X-ray source location at a height of 10 $R_G$ above a slowly-rotating (a = 0.1) black hole (red, dotted line) and a maximally spinning (a = 0.998) black hole (black, solid line).  {\it Right:} Lag vs frequency corresponding to the transfer functions shown in the left panel.}
\label{fig:spin}
\end{figure*}

We now consider the effect of black hole spin.  Figure~\ref{fig:spin} shows the transfer functions and corresponding lags for a slowly-rotating (a = 0.1) black hole and a maximally rotating (Kerr, a = 0.998) black hole.  Notice that the differences in the energy-averaged transfer function are subtle.  We discuss the differences when looking at the energy dependence of the lag in detail later \citep[see also the discussion in][comparing 2D transfer functions]{reynolds99}.  The most noticeable effect is a larger peak response for a maximally spinning black hole.  This is because for a maximally spinning black hole the disk has a larger surface area as it now extends closer to the black hole.  The response from the additional disk comes from the region with the strongest GR effects and thus a significant Shapiro delay, thus the additional response from having a disk that extends closer to the black hole does not occur straightaway (compared to the non-spinning black hole case), but after a short delay.  However, these small differences in the transfer function lead to similarly small differences in the frequency dependence of the lags.  Importantly, it has a negligible effect on the frequency where phase wrapping occurs.    For the maximally spinning black hole the extension of the inner disk to smaller radii means overall there will be more photons reflected (and less lost into the black hole), but the response from the outer part of the disk will still be the same, we therefore scaled the transfer functions in this case so that the response functions at late times match.

\subsection{Reflected response fraction}
\begin{figure}
\centering
\includegraphics[width=8cm]{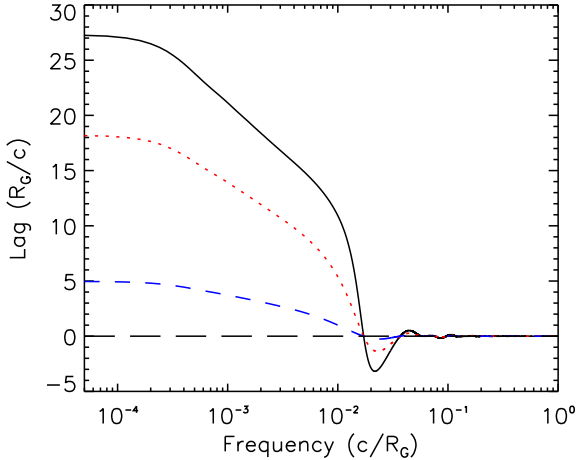}
\caption{The effect of changing the reflected response fraction.  Here we show the lag vs frequency for a maximally spinning black hole with and X-ray source located at $h = 10~R_G$ above the black hole, and $i = 45^\circ$.  The lines indicate $R = 1$ (solid, black), $R=0.5$ (dotted. red), $R=0.1$ (dashed, blue).}
\label{fig:refrac}
\end{figure}

We now consider the effect of changing the reflected response fraction.  The effect of having both the reflected component and the irradiating (power-law) component in the same observed band acts to dilute the observed lags. For instance, the lag between the power-law seen in both bands will be zero, and that will reduce the observed lag between the reflection and power-law component.  Such dilution effects are discussed in \citet{zog11_1h0707} and \citet{kara13a}, and \citet{wilkins13} discuss in more detail how it changes the lag vs frequency.  Here, we show an example of changing the reflected response fraction in Figure~\ref{fig:refrac}.   As can be clearly seen in Figure~\ref{fig:refrac}, changing the reflected response fraction acts to scale the lags.  The higher the reflected response fraction, the lower the dilution from the zero lag component, and hence the larger the measured lag.  We define $\bar{t}$ as the time corresponding to the mean value of the transfer function.  Then, using the definition that the reflected response fraction, $R$ = (reflection flux)/(power-law flux) the maximum measured lag (the value the function tends to at low frequencies) scales like $\bar{t}R/(1+R)$.  Hence, for a reflected response fraction of 1, the maximum lag will be 0.5$\bar{t}$, and for $R = 0.5$ the maximum lag will be $\bar{t}/3$.  This can be seen in Figure~\ref{fig:refrac}.   For the transfer function used there, $\bar{t} = 54~R_G/c$, and for $R=1$ we see a maximum lag of $27~R_G/c$, and for $R=0.5$ we see a maximum lag of $18~R_G/c$, as expected.  Note that here we have only considered the reflected response fraction across the Fe K band and do not consider the contribution from reflection in the band dominated by the irradiating flux.  However, this would only act to further dilute the lags in the same fashion \citep[for instance see figure 8 in][]{wilkins13}, and would be equivalent to reducing $R$. 

\subsection{Black hole mass}
\begin{figure}
\centering
\includegraphics[width=8cm]{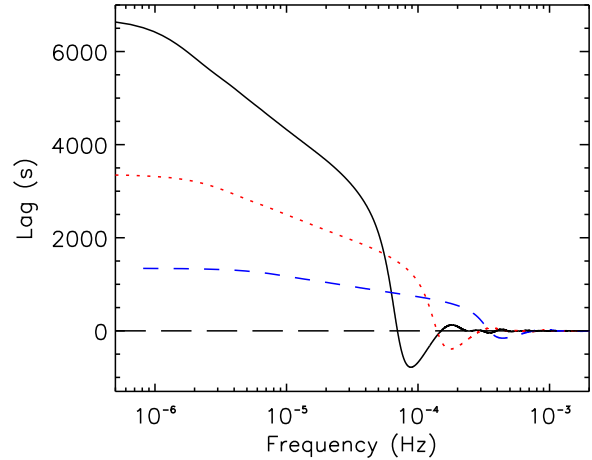}
\caption{The effect of including a specific black hole mass.  Here we show the lag vs frequency for a maximally spinning black hole with and X-ray source located at $h = 10~R_G$ above the black hole, and $i = 45^\circ$.  The lines indicate $M = 5\times10^7~M_\odot$ (solid, black), $M=2.5\times10^7~M_\odot$ (dotted, red), $M=10^7~M_\odot$ (dashed, blue).}
\label{fig:bhmass}
\end{figure}

Thus far, we have only shown the results in natural units including the gravitational radius, $R_G$.  Including a specific black hole mass would just scale all the times in the transfer function by a factor of $R_G/c = GM/c^3$.  This then has a corresponding effect on scaling the lags as $R_G/c$ and the frequencies as $c/R_G$.  We show the lag vs frequency for three different black hole masses in Figure~\ref{fig:bhmass}.

In summary, the main parameters which change the frequency dependence of the lag from the energy-averaged response is the height of the X-ray source above the black hole and the black hole mass.  The reflected response fraction is also important in scaling the absolute lags, while inclination and black hole spin do not have a strong effect.  However, the black hole mass and the height of the X-ray source are degenerate -- a larger black hole mass will increase the lag and decrease the frequency where phase wrapping occurs, while increasing the height of the X-ray source above the disk also has the same effect.  We have only assumed a simple lamppost geometry with a point source above the black hole, however, an extended corona (both radially and vertically) may be more realistic.  As explored by \citet{wilkins13}, having a vertically extended corona can change the frequency dependence of the lags, by changing both the average lag as well as the exact shape where phase-wrapping occurs.  In fact, it can lead to averaging-out of the negative lags (see their figure 4).

\subsection{Comparing with the lag frequency dependence of NGC 4151}

\citet{zoghbi12} calculated the lag between the 4 -- 5 keV and 2 -- 3 keV lightcurves as well as the 5 -- 6 keV and 2 -- 3 keV lightcurves.  The 2 -- 3 keV reference band was chosen as this is a region of the spectrum strongly dominated by the power-law component, while the 4 -- 5 keV and 5 -- 6 keV energy ranges sample different parts of the Fe K region.  Here we determine the frequency dependence of the lags in the 5 -- 6 keV bands as expected from the GR transfer functions discussed above, and compare with what was observed  in NGC~4151.  As before we assume no contribution from the reflected emission in the primary lightcurve.  As discussed above, changing this would just change the definition of $R$.

We assume an inclination of $45^\circ$ for NGC~4151, which is the inclination inferred based on detailed modeling of the narrow-line region in this object \citep{das05}.  Our discussion above shows how it is both the black hole mass and the height of the X-ray source above the black hole that are the two most important parameters for changing the frequency dependence of the lag, especially the frequency where phase-wrapping occurs and the lags tend to zero.  From the analysis of \citet{zoghbi12} we see that both the 4 -- 5 keV and 5 -- 6 keV lags are consistent with zero above a frequency of approximately $1.5\times10^{-4}$ Hz.  Similarly, the lag vs energy shows a clear Fe K line like profile for the frequency range $(1 - 2)\times10^{-5}$ Hz, with lags in the 5 -- 6 keV band of approximately 2000 seconds.  Both these facts already allow constraints on the black hole mass and height of the X-ray source.

\begin{figure}
\centering
\includegraphics[width=8cm]{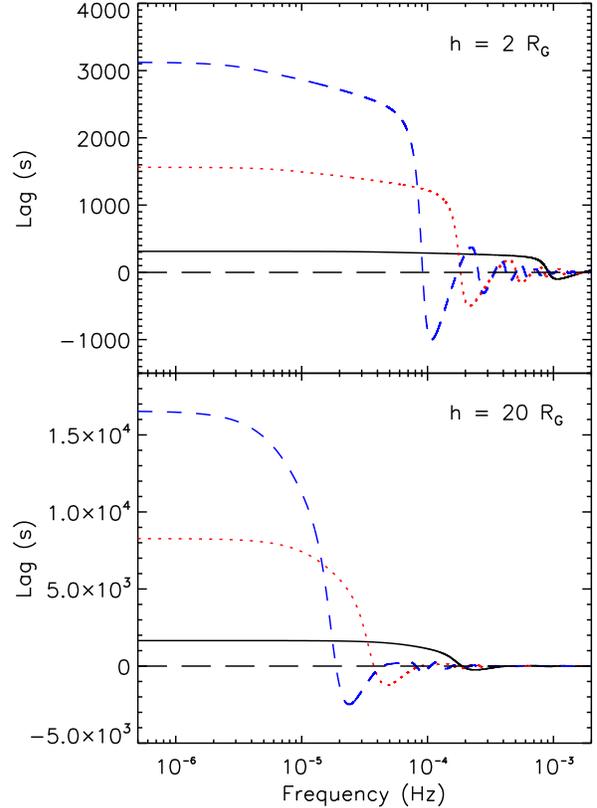}
\caption{Lag vs frequency for the 5 -- 6 keV band, assuming a reflected response fraction of 1 in that band.  The different lines assume different black hole masses of $10^7~M_\odot$ (solid, black), $5\times10^7~M_\odot$ (dotted, red), $10^8~M_\odot$ (dashed, blue).  The top panel has the X-ray source located at $h = 2~R_G$ above the black hole, while the bottom panel has it at $h = 20~R_G$.}
\label{fig:lagvfreq}
\end{figure}

In Figure~\ref{fig:lagvfreq} we show several models for lag versus frequency using the average response in the 5 -- 6 keV band, and assuming a reflected response fraction of 1 in that band.  In the top panel we show the lag vs frequency for $h = 2~R_G$, while in the bottom panel we show it for $h = 20~R_G$.  The different lines show different masses of 1, 5 and 10 $\times10^7~M_\odot$.  We can already conclude several things about $h$ and $M$ based on this.  Firstly, the fact that we observe lags in the 5 -- 6 keV (and 4 -- 5 keV) range up to about $1.5\times10^{-4}$ Hz rules out models with $h = 20~R_G$ and a black hole mass much greater than $10^7~M_\odot$.  For $h=2~R_G$ a mass of $10^8~M_\odot$ turns negative at $10^{-4}$ Hz, again not what is observed.  Models with lower masses do lead to positive lags above this frequency, but, the lags are short and would require a larger than realistic reflected response fraction for a mass of $10^7~M_\odot$.  Given that the optical reverberation mapping mass for NGC~4151 of $4.6\times10^7~M_\odot$ (similar to the dotted line in Figure~\ref{fig:lagvfreq}) this suggests $h$ greater than 2 but less than 20 $R_G$.  The degeneracy between black hole mass and $h$ is quite apparent in Figure~\ref{fig:lagvfreq} and a higher black hole mass requires a smaller $h$.

\begin{figure}
\centering
\includegraphics[width=8cm]{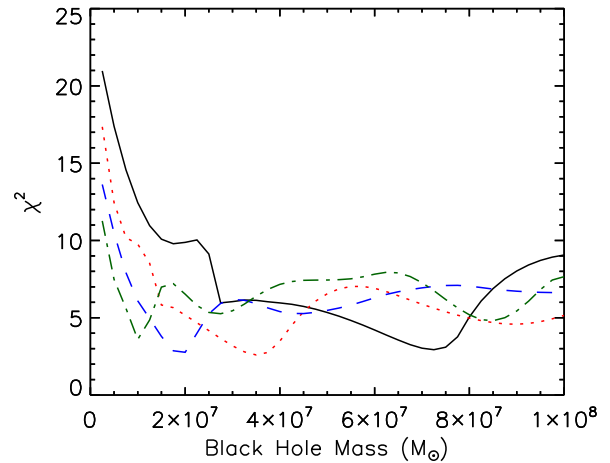}
\caption{The $\chi^2$ distribution with black hole mass when fitting the lag vs frequency data for the 5 -- 6 keV band for NGC 4151 from \citet{zoghbi12}.  The different lines assume a different height of the X-ray source with $h = 2~R_G$ (solid, black), 5$~R_G$ (dotted, red), 10$~R_G$ (dashed, blue), 20$~R_G$ (dot-dashed, green).  The reflected response fraction was a free parameter in the fit. Notice how the best fit black hole mass decreases with increasing $h$.}
\label{fig:chi2}
\end{figure}

\begin{figure}
\centering
\includegraphics[width=8.5cm]{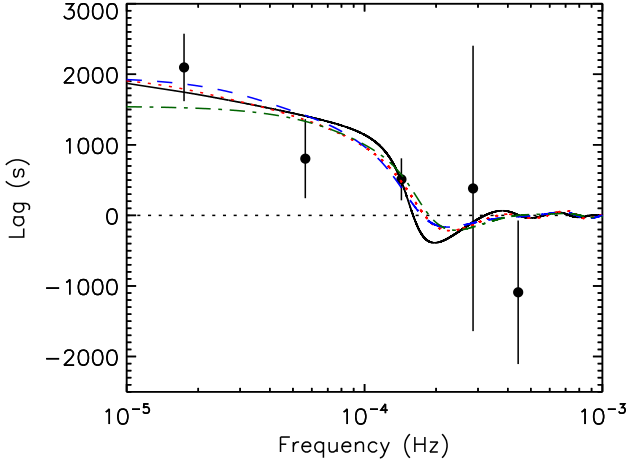}
\caption{Data points show the lag vs frequency for the 5 -- 6 keV band for NGC 4151 from \citet{zoghbi12}.  The different lines assume a different height of the X-ray source with $h = 2~R_G$ (solid, black), 5$~R_G$ (dotted, red), 10$~R_G$ (dashed, blue), 20$~R_G$ (dot-dashed, green), and the best-fitting black hole mass for each $h$ is used.}
\label{fig:lagvfreq_fit}
\end{figure}

We now fit the black hole mass and reflected response fraction needed for each $h$ to the 5 -- 6 keV band lag vs frequency.   For each value of the black hole mass we determine the reflected response fraction that minimizes $\chi^2$.  In Figure~\ref{fig:chi2} we show $\chi^2$ as a function of black hole mass for $h = 2, 5, 10, 20~R_G$.  Note how the $\chi^2$ minimum is at lower black hole masses for larger values of $h$.  The overall  minimum $\chi^2$ value is 2.59 (for three degrees of freedom: 5 data points and two parameters) and corresponds to $h = 5~R_G$ and $M = 3.5\times10^7~M_\odot$, but the minimum values for other values of $h$ are similarly low.  We show the best-fitting lag vs frequency for the 4 values of $h$ in Figure~\ref{fig:lagvfreq_fit}.  The best fitting mass for $h = 5~R_G$ corresponds to $M = 3.5\times10^7~M_\odot$ and the best fitting mass for $h = 2~R_G$ corresponds to about $M = 7.5\times10^7~M_\odot$.  The corresponding reflected response fractions are $R = 0.92$ for $h = 5~R_G$ and $R = 0.91$ for $h = 2~R_G$.  Considering that the optical reverberation mapping mass for the black hole in NGC~4151 is $4.6\times10^7~M_\odot$, our results based on the frequency dependence alone, would therefore imply that the height of the X-ray source is somewhere between $2 - 5 R_G$.

We note that the secondary dips in $\chi^2$ that can be seen at higher masses in Figure~\ref{fig:chi2} occur due to the oscillatory nature of phase wrapping and a minimum occurs when the peaks match with the data.  This is unavoidable when considering just the frequency dependence.  In the next section we consider the energy dependence on the lag, which rules out these secondary solutions.

Our initial approach here has been to look at the simplest lamppost geometry model.  As we note earlier, vertically and/or radially extended corona models may be more realistic and an initial investigation of the frequency-dependence of such models was presented by \citet{wilkins13}.  Extended  corona models are beyond the scope of this work we note that extending the corona radially shifts the phase wrapping frequency to lower values while also lowering the average lag.  This can be seen in figure 4 from \citet{wilkins13}.  To counteract the phase wrapping occurring at lower values would require a lower mass black hole or height for the corona.  But, such changes in mass and height would also reduce the average lag further.  Increasing the average lag would require a higher reflected response fraction.  We expect the changes in mass, height and reflection fraction to be relatively small, though the changes will be larger for a more extended corona.  We will explore these dependencies and the differences with lamppost models in more detail in future work.

\section{Energy dependence of the lag}\label{sec:enevo}

The energy dependence of the lags is also an important consideration, especially given that is where we see the signature of lags in the Fe K line.  In this section we start by considering how inclination, height of the X-ray source, spin, black hole mass and observed frequency range set the lags as a function of energy.  We then go on to compare the observed Fe K lags in NGC~4151 with these models.

To calculate the energy-dependent lags we use the 2D GR transfer functions described above.  Now, instead of averaging over the entire iron line, we look at the response of the disk at every energy.  For the purposes of illustrating the general properties, we normalize the entire transfer function so that the reflected response fraction at the peak of the Fe K$\alpha$ line is 1 (in other words the transfer function is normalized so that the area under the transfer function at the energy where the line peaks is 1).  We now go through a few examples to show how the lags change as a function of energy for different parameters.

\subsection{Lag as a function of energy and frequency}
\begin{figure*}
\centering
\includegraphics[width=16cm]{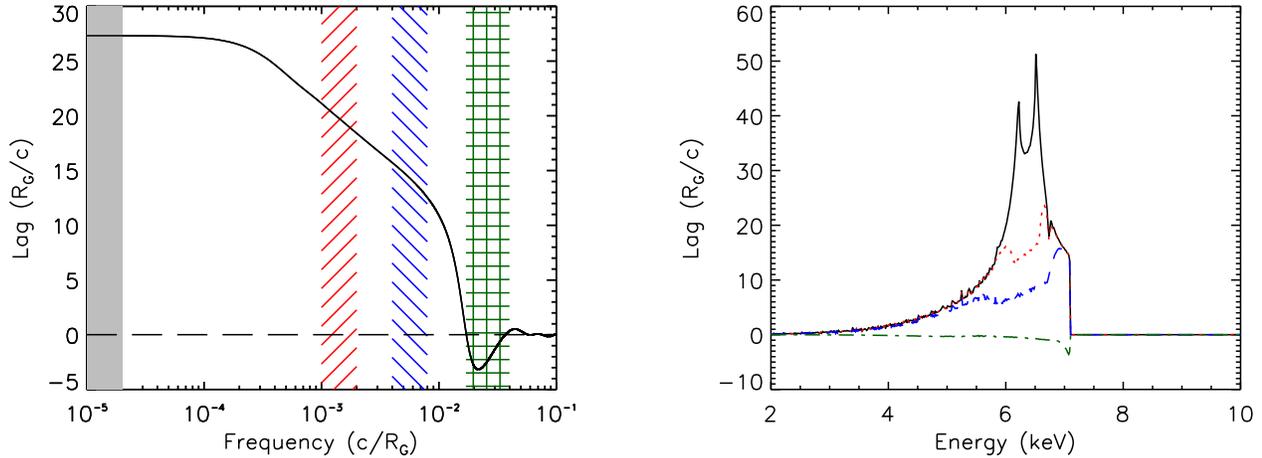}
\caption{{\it Left:} Lag vs. frequency for the entire Fe K$\alpha$ line.  The colored regions correspond to the frequency ranges used to calculate the lags as a function of energy. {\it Right:} Lag as a function of energy for four different frequency ranges: $(1 - 2)\times10^{-5}~c/R_G$ (black, solid line), $(1 - 2)\times10^{-3}~c/R_G$ (red, dotted line), $(4 - 8)\times10^{-3}~c/R_G$ (blue, dashed line) and $(1.7 - 4)\times10^{-2}~c/R_G$ (green, dot-dashed line).  These are all calculated for $h = 10~R_G$, $i = 45^\circ$ and a maximally spinning black hole.}
\label{fig:tf_lagspec_plot}
\end{figure*}

When looking at the energy-dependence of the lag, we choose a frequency range to examine.  The energy-dependence also has a dependence on the choice of frequency range.  To start with, we show the energy-dependence of the lag for several different frequency ranges, in Figure~\ref{fig:tf_lagspec_plot}.  We choose $h = 10~R_G$, $i = 45^\circ$, and $a = 0.998$.  Figure~\ref{fig:tf_lagspec_plot} shows that on the longest timescales (lowest frequencies) we see the lags from the entire disk (black solid line, right panel).  Particularly prominent are the large lags showing a double-horned profile from the outer disk.  Of course, we also see the broadened red-wing due to the response from the inner disk.  As we go to higher frequency ranges (red dotted and blue dashed lines) we see a cut-out region in the middle of the lag profile.  This is due to the fact that at these higher frequencies we do not see a response from the outer disk.  This cut-out region gets broader at higher frequencies (compare the red dotted and blue dashed lines) as we see less and less response from the outer parts of the disk.  Finally, once we get to the frequency range where the phase wraps completely, with the lags going negative (green dot-dashed line) we see very small, negative lags showing a mirror-image of the Fe K$\alpha$ lags at lower frequencies.  Thus, picking out different frequency ranges probes different parts of the disk.  At the lowest frequencies we see the entire disk, while as we move to higher frequencies we progressively start to filter out all but the response from the inner disk, before we end up with zero lags.  Note that here we have used relatively narrow frequency ranges for this example.  A broader frequency range such as $(1 - 8)\times10^{-3}~c/R_G$ would give a lag profile that is approximately an average of the red dotted and the blue dashed lines.

Now we have examined the general behaviour as a function of energy and frequency for one GR transfer function, we will explore how each of the parameters effects the lag profile.

\subsection{Height of the X-ray source}
\begin{figure*}
\centering
\includegraphics[width=16cm]{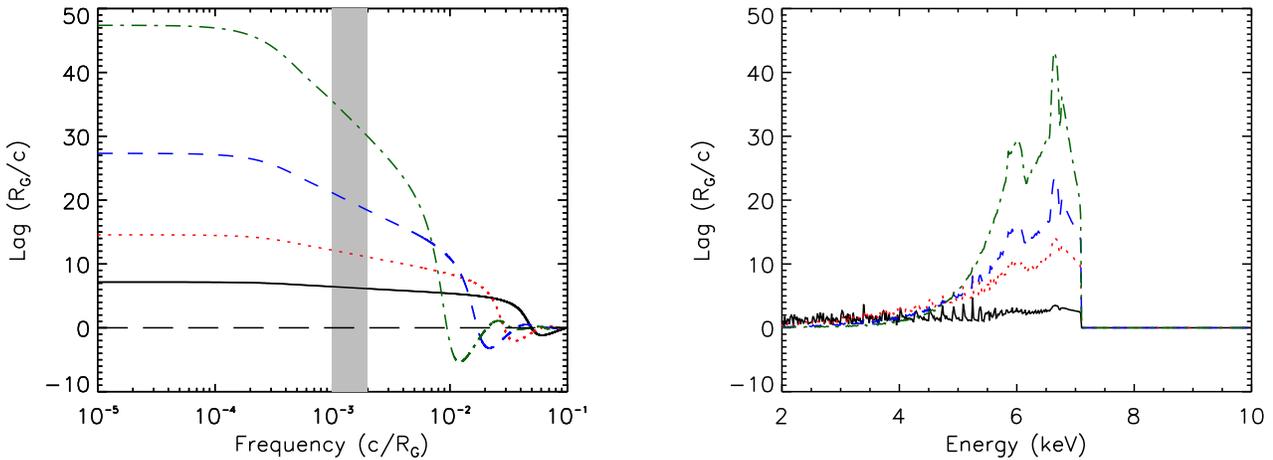}
\caption{The effects of the height of the X-ray source on lags.  {\it Left:} Lag vs. frequency for the entire Fe K$\alpha$ line for heights of $h = 2~R_G$ (black, solid), $5~R_G$ (red, dotted), $10~R_G$ (blue, dashed) and $20~R_G$ (green, dash-dotted).  The gray box shows the frequency range used for the lag vs energy plots in the right panel.  {\it Right:} Lag vs energy in the frequency range $(1 - 2)\times10^{-3} c/R_G$ for heights corresponding to the lines in the left panel. These are all calculated for $i = 45^\circ$ and a maximally spinning black hole.}
\label{fig:tf_lagspec_h}
\end{figure*}

In  Figure~\ref{fig:tf_lagspec_h} we show how the height of the X-ray source changes the observed lags as a function of energy in a given frequency range. As with the energy-averaged lags discussed earlier, an increase in $h$ leads to longer lags.  Another change is that when the X-ray source is very close to the black hole ($\leq 5~R_G$) the response from the inner disk becomes stronger and hence the lags in the red wing appear longer as they are less diluted by the continuum, though this is a subtle effect.  One can see that the height does not affect the energies where the cut-out occurs, or the general shape of the lag profile (spikes for $h=2~R_G$ are numerical in origin).

\subsection{Inclination}

\begin{figure*}
\centering
\includegraphics[width=16cm]{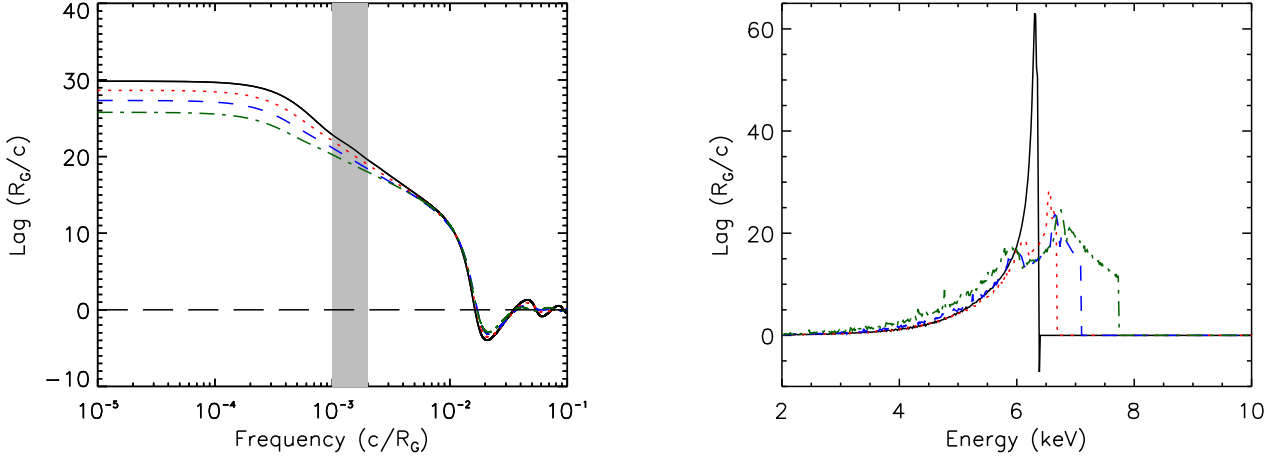}
\caption{The effects of inclination of the disk on lags.  {\it Left:} Lag vs. frequency for the entire Fe K$\alpha$ line for inclinations of $i = 5^\circ$ (solid, black), $30^\circ$ (dotted, red), $45^\circ$ (dashed, blue) and $60^\circ$ (dash-dotted, green).  The gray box shows the frequency range used for the lag vs energy plots in the right panel.  {\it Right:} Lag vs energy in the frequency range $(1 - 2)\times10^{-3} c/R_G$ for inclinations corresponding to the left panel. These are all calculated for $h = 10~R_G$ and a maximally spinning black hole.}
\label{fig:tf_lagspec_i}
\end{figure*}

Inclination only has a small effect on the frequency dependence of the lags, but, there is a much more significant effect when looking at lags as a function of energy.  As with the time-averaged response of the accretion disk \citep[e.g.][]{fabian89}, the inclination has an significant influence on the extent of the blue wing of the line.  As the inclination increases (goes from face-on to edge-on), the component of the velocity along the line of sight increases, and thus Doppler shifts increase.  This is seen most prominently in the energy of the blue wing of the iron line.  The same effects shape the energy where we see lags, thus, higher inclinations have lags to higher energies and the narrowest lag profile comes from the lowest inclination (see Figure~\ref{fig:tf_lagspec_i}).

\subsection{Black Hole Spin}

\begin{figure*}
\centering
\includegraphics[width=16cm]{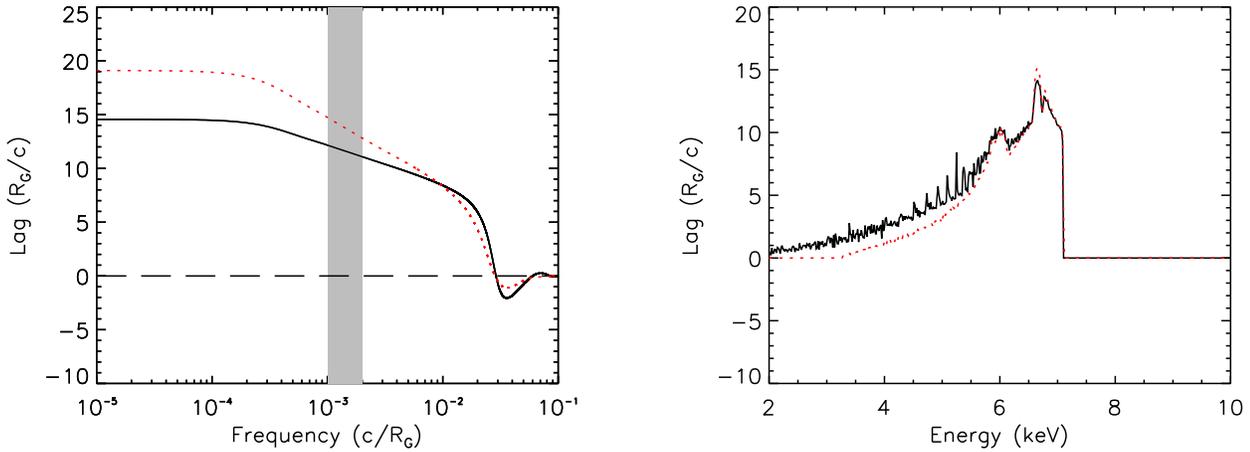}
\caption{The effects of black hole spin on lags.  {\it Left:} Lag vs. frequency for the entire Fe K$\alpha$ line for black hole spins of $a = 0.1$ (red, dotted line) and 0.998 (black, solid line).  The gray box shows the frequency range used for the lag vs energy plots in the right panel.  {\it Right:} Lag vs energy in the frequency range $(1 - 2)\times10^{-3} c/R_G$ for $a = 0.1$ (black, solid line) and $a = 0.998$ (red, dotted line) These are all calculated for $h = 5~R_G$ and $i = 45^\circ$.}
\label{fig:tf_lagspec_a}
\end{figure*}

Next, we compare the lag energy profiles for different black hole spins.  In order to highlight the difference in the lags with spin, we use $h = 5~R_G$.  As with the time-averaged iron line, an increase in the spin of the black hole broadens the red-wing of the line.  So, here we see a more prominent red-wing of the profile as more emission is coming from the innermost radii as well as the lags extending to lower energies for a higher spin (see~Figure~\ref{fig:tf_lagspec_a}).  The effect is quite subtle, and would require very accurate measurements to be detectable.

\subsection{Black Hole Mass}

\begin{figure*}
\centering
\includegraphics[width=16cm]{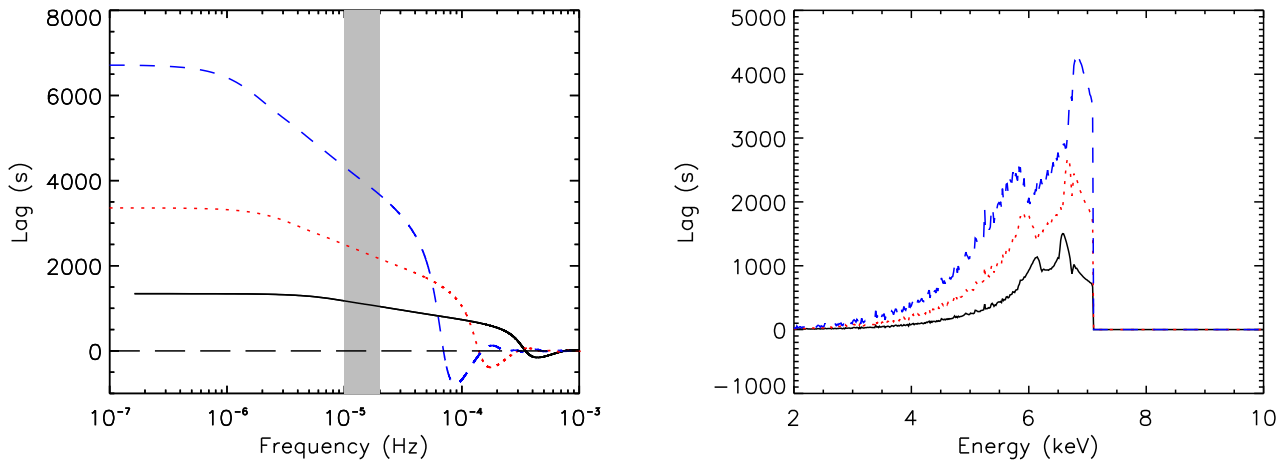}
\caption{The effects of black hole mass on lags.  {\it Left:} Lag vs. frequency for the entire Fe K$\alpha$ line for black hole masses of $10^7~M_\odot$ (black, solid line), $2.5\times10^7~M_\odot$ (red, dotted) and $5\times10^7~M_\odot$ (blue, dashed line).  The gray box shows the frequency range used for the lag vs energy plots in the right panel.  {\it Right:} Lag vs energy in the frequency range $(1 - 2)\times10^{-5}$ Hz for $10^7~M_\odot$ (black, solid line), $2.5\times10^7~M_\odot$ (red, dotted) and $5\times10^7~M_\odot$ (blue, dashed line). These are all calculated for $h = 10~R_G$, $i = 45^\circ$, and $a = 0.998$.}
\label{fig:tf_lagspec_mass}
\end{figure*}

Finally, we look at how a different black hole mass changes the lags as a function of energy in a given frequency range.  In Figure~\ref{fig:tf_lagspec_mass} we show the lag profile in the frequency range $(1 - 2)\times10^{-5}$ Hz for black hole masses of $10^7~M_\odot$, $2.5\times10^7~M_\odot$ and $5\times10^7~M_\odot$ (all assuming $h = 10~R_G$, $i = 45^\circ$, and $a = 0.998$).  As can be seen from the frequency dependence of the energy averaged lags, the more massive the black hole, the larger the lags.  Also, the mass of the black hole shifts the frequencies at which phase wrapping occurs.  This means that for a fixed frequency range, with a lower mass black hole we will be looking at lags nearer to the maximum lag than for a more massive black hole.  Both these effects can be seen in the lag-energy profile -- lags from a more massive black hole are longer, but also, given the fixed frequency range the lags from the most massive black hole have a larger cut-out region, mostly shows lags from the inner disk.  On the other hand, the lowest mass black hole shows no cut-out region, and lags from almost the entire disk.  While the height of the X-ray source and the black hole mass are highly degenerate when looking at the lag vs frequency, they are also somewhat degenerate when looking at the lag vs energy.  However, the fact that as the mass changes the fraction of the disk that we see a response from on that frequency changes can help break this degeneracy in the case of very good data.

\subsection{Comparison with NGC~4151}
We now compare the observed lags in NGC~4151 with the models presented above.  The lags in NGC~4151 show lags of a few thousand seconds around the Fe K$\alpha$ line energies \citep{zoghbi12}.  However, note that the lags abruptly drop to zero above 6.5 keV.  Looking at the lags as a function of energy presented above, for the case of $i = 45^\circ$, we actually expect the lags to be significant to above 7 keV.  While we note that there is an extremely strong, narrow and constant Fe K$\alpha$ line seen in the spectrum of NGC~4151 in addition to a broad Fe~K$\alpha$ line, a constant component like this should not affect the lags measured (only variable components contribute to lag measurements).

\begin{figure}
\centering
\includegraphics[angle=270,width=8.4cm]{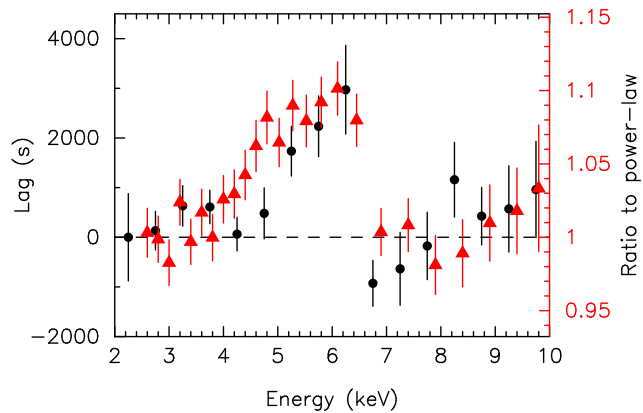}
\caption{Black circles show the lags as a function of energy for NGC 4151 in the frequency range $f<2\times10^{-5}$ Hz \citep[data from][]{zoghbi12}.  Red triangles show the ratio of the difference spectrum of NGC 4151 to a power-law (fit ignoring the energy range 4 -- 8 keV).  The difference spectrum shows that the Fe~K$\alpha$ line is not variable above approximately 6.5 keV.  Correspondingly, the lags drop significantly at the same energy.}
\label{fig:diffspec}
\end{figure}

We look to see at which energies the Fe~K$\alpha$ line is variable by studying the difference spectrum between the low and high states used in \citet{zoghbi12} (see their figure 3 for the spectra).  As discussed in  \citet{zoghbi12} the difference spectrum is not a simple power-law and has residuals that resemble a broad iron line when fitted with a power-law ignoring the 4 -- 8 keV region.  As can be seen in Fig.~\ref{fig:diffspec} both the difference spectrum and the lags match each other in shape closely, especially an abrupt drop at around 6.5 keV.  The drop in the difference spectrum indicates the Fe~K$\alpha$ line is not variable above $~$6.5 keV.  The inclination in our model, then, must not gives lags above this energy.  This means we expect to get the best fits from low inclinations ($i < 30^\circ$), as the blue wing of the line will shift below 6.5 keV for those inclinations.  Higher inclinations will give non-zero lags above 6.5 keV.

As we showed above, when looking at the frequency dependence of the lags in the 5 -- 6 keV band, a range of masses are consistent with the observed evolution and that mass and source height are degenerate.  Here, when comparing the lag as a function of energy we choose to fix the black hole mass to be the optical reverberation mapping mass of $4.6\times10^7~M_\odot$.  We allow the reflected response fraction at the peak of the line to be a free parameter, and explore fitting the lags with each value of $h$,  $i$, and $a$.  We calculate a $\chi^2$ value for the fit using only the data points below 7.5 keV given that above 7.5 keV all models give zero lag.  

In Table~\ref{tab:lagenfit} we give the best-fitting reflected response fraction (at the peak of the line), and the $\chi^2$ value for each combination of $h$, $i$,  and $a$.  As can be seen, good fits (reduced-$\chi^2 \sim 1$) are only found for inclinations less than 30$^\circ$, as expected.  We find the best fits for $h = 5~R_G$, and note that the fits gets significantly worse for $h = 2~R_G$.  The black hole spin does not have a large effect on the quality of the fits, though we find that $a = 0.1$ typically has a $\chi^2$ value approximately 1 higher than the fits with the same $h$ and $i$ but $a = 0.998$, in other words we find $a=0.998$ is a better fit at approximately the 1$\sigma$ level.  The best-fits have a reflected response fraction at the peak of the line of approximately 1.

\begin{figure}
\centering
\includegraphics[width=8.4cm]{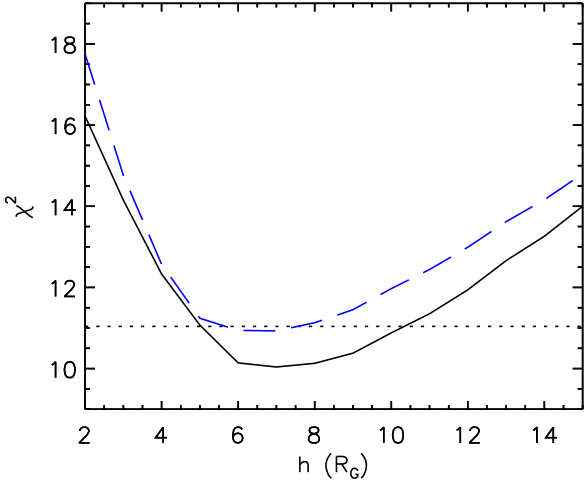}
\caption{$\chi^2$ as a function of source height, $h$, for $i = 5$ (black, solid line) and $i = 20$ (blue, dashed line) and $a=0.998$ in both cases. The dotted line indicates $\Delta\chi^2 = 1.0$, i.e. the $1\sigma$ confidence level.}
\label{fig:chi2h}
\end{figure}

Given that the best fits are for $i = 5^\circ$ and $i =20^\circ$ with $h = 5~R_G$ and $a = 0.998$, we chose to further explore the dependence of the fits on $h$ for these two inclinations from $h = 2$ to $h = 15~R_G$.  We find the global best-fit for $i = 5^\circ$ at $h = 7~R_G$, with $1\sigma$ range ($\Delta\chi^2 = 1.0$) from 4.5 to 9.9 $R_G$.  The best-fit for $i=20^\circ$ is also at $h = 7~R_G$, with a $\chi^2$ value within $1\sigma$ of the global best-fit.  We show $\chi^2$ as a function of $h$ in Figure~\ref{fig:chi2h}.  The reflected response fractions for these best fits are $R = 1.05$ and $R = 1.11$ for $i = 5^\circ$ and $i=20^\circ$, respectively.

\begin{figure}
\centering
\includegraphics[width=8.4cm]{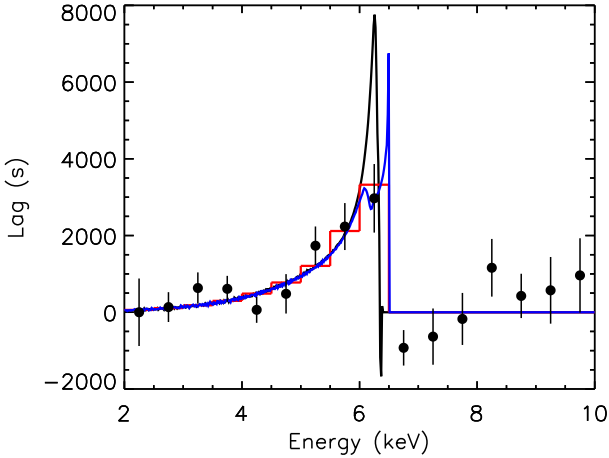}
\caption{Lags in NGC 4151 in the frequency range $(1 - 2)\times10^{-5}$~Hz (black circles) along with the best-fitting models.  The models assume the optical reverberation mapping mass for NGC 4151 ($4.6\times10^7 M_\odot$) and are calculated for $h = 7~R_G$, with $i = 5^\circ$ (black) and $i=20^\circ$ (blue).  The red line shows the average for the $i = 5^\circ$ model with the same energy binning as the data.  The binned $i=20^\circ$ is virtually identical and thus is not shown for sake of clarity. }
\label{fig:lagspecbestfit}
\end{figure}

\begin{figure}
\centering
\includegraphics[width=8.4cm]{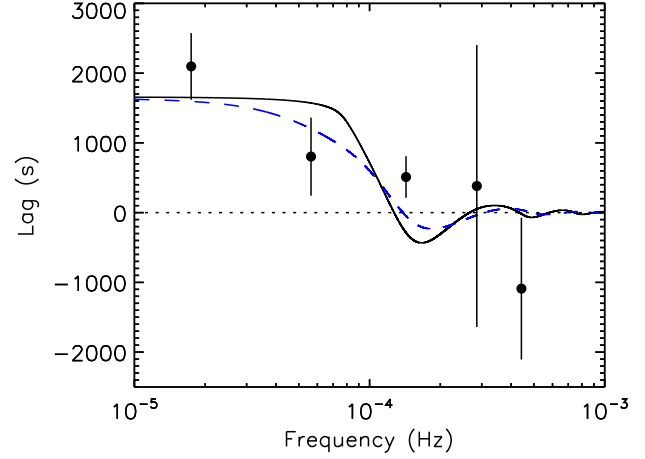}
\caption{Frequency dependence of lags in the 5 -- 6 keV range in NGC 4151 (black circles).  The models shown are those found from fitting the energy dependence of the lags in the $(1-2)\times10^{-5}$ Hz range (the frequency dependence has not been simultaneously fit). The models assume the optical reverberation mapping mass for NGC 4151 ($4.6\times10^7 M_\odot$) and are calculated for $h = 7~R_G$, with $i = 5^\circ$ (black, solid) and $i=20^\circ$ (blue, dashed).}
\label{fig:lagfreqbestfit}
\end{figure}

We show the best-fitting models for a maximally spinning black hole with $h = 7~R_G$ and  $i=5^\circ$ and $i=20^\circ$ in Figure~\ref{fig:lagspecbestfit}.  We also show the frequency dependence of the lags for these best-fitting models compared to NGC 4151 in Figure~\ref{fig:lagfreqbestfit}, though note that we do not simultaneously fit the frequency and energy dependence of the lags here.  They are all broadly consistent with the data, and higher signal-to-noise ratio measurements would be needed to distinguish between them.
 
It is interesting to note that the best-fitting models for $i=5^\circ$ and $i=20^\circ$ are almost identical apart from in the $6.0 - 6.5$ keV energy bin.  While the average value across that energy band is also virtually the same, the $i=5^\circ$ peaks at larger lags yet goes to zero at a lower energy.  Clearly with better data there would be the possibility of better constraining the inclination from fitting the lags as a function of energy.

\begin{table}
\caption{Parameters for fitting the lag vs. energy in the frequency range $(1-2)\times10^{-5}$ Hz in NGC 4151. For each $h$, $i$, and $a$ the best-fitting reflected response fraction, $R$, was found.  The degrees of freedom are therefore $\nu = 10$ based on the 11 data points below 7.5 keV where $\chi^2$ evaluated and the 1 free parameter in the fits.}
\centering
\begin{tabular}{ccccc}
\hline
$h$ & $i$ & $a$ & $R$ & $\chi^2$ \\
\hline
2 & 5 & 0.998 & 0.77  & 16.2 \\
2 & 20 & 0.998 & 0.84  & 17.7 \\
2 & 30 & 0.998 &  0.67 & 28.6 \\
2 & 45 & 0.998 &1.25 & 36.7 \\
5 & 5 & 0.1 & 0.86  & 12.0 \\
5 & 20 & 0.1 & 1.20 & 13.3 \\
5 & 30 & 0.1 & 0.68 & 30.0 \\
5 & 45 & 0.1 & 0.17 & 46.3 \\
5 & 5 & 0.998 & 0.66  & 11.1 \\
5 & 20 & 0.998 &  1.00 & 11.2 \\
5 & 30 & 0.998 & 0.58 & 25.9 \\
5 & 45 & 0.998 & 0.27 & 42.4 \\
10 & 5 & 0.1 & 0.82 & 12.0 \\
10 & 20 & 0.1 & 1.13 & 13.9 \\
10 & 30 & 0.1 & 0.60 & 32.9 \\
10 & 45 & 0.1 & 0.10 & 46.4 \\
10 & 5 & 0.998 & 1.15  & 10.9  \\
10 & 20 & 0.998 & 1.18 & 12.0 \\
10 & 30 & 0.998 & 0.61  & 30.6 \\
10 & 45 & 0.998 & 0.13 & 45.5 \\
20 & 5 & 0.1 & 0.98 & 16.8 \\
20 & 20 & 0.1 & 1.14  & 18.1 \\
20 & 30 & 0.1 &  0.43 & 38.1 \\
20 & 45 & 0.1 & 0.06  & 46.6 \\
20 & 5 & 0.998 & 0.97  & 16.2 \\
20 & 20 & 0.998 & 1.14  & 17.4 \\
20 & 30 & 0.998 & 0.44 & 37.5 \\
20 & 45 & 0.998 &  0.07 & 46.4 \\

\hline
\end{tabular}
\label{tab:lagenfit}
\end{table}

Looking at the higher frequency range used in \citet{zoghbi12} of $(5-50)\times10^{-5}$ Hz we find that all the best-fitting models to the $(1-2)\times10^{-5}$ Hz frequency range give lags that are close to zero (less than a few hundred seconds).  The size of the frequency range means that for all models it covers some of the range where the lags go to zero.  The lags calculated from the NGC~4151 data are measured relative to the reference band lightcurve and thus the exact zero-point of the lags is not known.  However, a straight line can fit the $(5-50)\times10^{-5}$ Hz lags reasonably well, which could be consistent with near-zero lags across all energies.

We note that in fitting the frequency dependence of the lags we assumed $i=45^\circ$, while the energy dependence of the lags suggests a significantly lower inclination than this.  However, the choice of inclination does not significantly affect the lags over the 5 -- 6 keV band.  As we show in Figure~\ref{fig:inc}, the energy-averaged response is not highly dependent on the choice of inclination.  Furthermore, as we show in Figure~\ref{fig:lagfreqbestfit}, the best fits to the energy dependent lags are also broadly consistent with the frequency dependence of the lags.

Spectral fitting of the time-averaged energy spectrum leads to a reflected response fraction at the peak of the line of about 0.8 \citep[see figure 3 in][]{zoghbi12}.  Our best fitting values from fitting the lag energy dependence gives $R \sim 1$.   This slightly higher value can be accounted for by noting that it is the relative strengths of the variable components of the power-law and reflection emission that matter. The time-averaged spectrum is not necessarily measuring this as one or both of these emission components will have constant, non-variable part.

It is also important to note that our model does not contain any continuum lags (often referred to has `hard lags') that are often seen at the frequencies where we observe the Fe~K$\alpha$ lag in NGC~4151.  These hard lags usually manifest themselves as a monotonic increase from low to high energies \citep[see figure 2 in][for instance]{kara13a}.  But, we note that the lags at 2 -- 3 keV (where contribution from the Fe~K$\alpha$ line is negligible) are consistent with the lags at energies above 7 keV (where the Fe~K$\alpha$ line does not contribute), indicating that there are no significant hard continuum lags seen in NGC~4151 at these frequencies.

In summary, given the optical reverberation mapping mass, an X-ray source with height of $7~R_G$ and a disk inclination of $<30^\circ$ can reproduce the observed energy dependence of the lags in NGC 4151.

\section{Conclusions}\label{sec:discuss}

We have used general relativistic transfer functions to demonstrate the basic properties of the lags expected from a simple lamppost type geometry where an X-ray point source is located some height $h$ above the black hole.  We have explored the dependence of the lags on both frequency and energy.  The two most important parameters that set the frequency dependence of the Fe K$\alpha$ lags is the black hole mass and the height of the X-ray source.  These two parameters are highly degenerate -- the more massive the black hole, the longer the lags, but similarly the larger the value of $h$ the longer the lags.  These parameters also set the frequency at which phase wrapping occurs and the lags oscillate around zero.  Source inclincation and black hole spin have a much smaller effect.  The reflected response fraction also plays a key role in the observed lags, as emission from the direct component in addition to emission from the reflected component acts to dilute the observed lag.  This dilution effect has been discussed in detail previously \citep{zog11_1h0707, kara13a, wilkins13}.

The energy-dependence of the lags over the Fe K$\alpha$ line region also depend highly on black hole mass and the height of the X-ray source, and once again, these parameters are degenerate.  The effects of black hole spin and inclination are much more important in shaping the lag profile than they are when looking at the frequency dependence.  As with the time-averaged Fe K$\alpha$ line, higher inclination sources increase the energy of the blue wing, here too, when looking at the lags, the energy where the blue wing of the lag profile occurs also increases with increasing energy.  For a higher black hole spin more emission comes from the innermost regions as so we see longer lags in the red wing of the lag profile, and hence, there is potential to measure the black hole spin through sensitive measurements of Fe K$\alpha$ lags.  

The frequency range over which the lag profile is determined  is also important (see Figure~\ref{fig:tf_lagspec_plot}).  At the lowest frequencies the response from the entire disk is observed, resulting in a lag profile with a double-horned shape.  As one moves to higher frequencies, the response from the outer parts of the disk is not seen, leading to a `cut-out' region in the lag profile.  This `cut-out' region becomes broader with increasing frequency as we move to seeing only the response from the inner disk with the shortest lags.  At the highest frequencies the lags oscillate around zero.

We compared both the lag vs. frequency and lag vs. energy observed for NGC~4151 with the lags calculated from the general relativistic transfer functions.  As a first step, we took the simplest approach of looking at the frequency and energy dependence separately.  Our models tested $h = 2, 5, 10$ and $20~R_G$.  The frequency dependence of the lags in the 5 -- 6 keV range can be explained by any of these heights if the black hole mass is left as a free parameter in the fit.  $h=2~R_G$ gives a best fit with a mass of $7.5\times10^7~M_\odot$ while $h = 20~R_G$ gives a best fit with a mass of $1\times10^7~M_\odot$.  The optical reverberation mapping mass for the black hole in NGC~4151 is $4.6\times10^7~M_\odot$, which, based on the frequency dependence alone would imply $h$ close to $5~R_G$.

When looking at the lags as a function of energy we assume the optical reverberation mapping mass for the black hole given the degeneracy between $h$ and black hole mass.  We fit the lags across the Fe K$\alpha$ line region in the frequency range $(1-2)\times10^{-5}$~Hz, allowing the reflected response fraction to be a free parameter in the fit.  We get a best fit for $h = 7~R_G$ and a $1\sigma$ confidence interval of $4.5 - 9.9~R_G$.  The fact that the lags drop to zero above 6.5 keV implies an inclination of $<30^\circ$, which suggests that the inner disk is not aligned with the narrow line region, and would also seem at odds with the Seyfert 1.5 classification for this source, which might imply the disk and broad line region are mis-aligned.  However, better constraints on the inclination would require much finer energy resolution in the lags in the 6.0 -- 6.5 keV energy range. We find that a maximally spinning black hole gives a slightly improved fit (at the $1\sigma$ level) compared to a black hole with $a = 0.1$, and clearly this method could be used to measure black hole spin if highly constrained lags in the red wing of the line could be achieved. 

Naively taking the longest lag in NGC 4151 and converting it to a size scale (by dividing by $c$) would give an average (emissivity-weighted) distance between the X-ray source and the reflecting region of $13~R_G$.  Our approach of fitting the lags with a general relativistic transfer function and fitting for dilution effects also gives a similar height of $7~R_G$ for a black hole mass of $4.6\times10^7~M_\odot$, and matches both the energy and frequency dependence of the lags.  The observed lags are completely consistent with this optical reverberation mass. Note that from the frequency dependence of the lags alone, a black hole more massive than approximately $8\times10^7 M_\odot$ would be hard to explain as it would require that the height of the X-ray source was less than $2~R_G$.  Furthermore, the energy dependence of the lags is also inconsistent with a X-ray source that close to the black hole as it would lead to too broad a lag profile.

The dependence of the lags on both frequency and energy provides a powerful way to determine the geometry of the reflection region.  Here, we have taken the simplest approach of fitting only the frequency or energy dependence of the lags separately when comparing to NGC~4151, yet, in the future with much more sensitive measurements in the Fe K$\alpha$ region, we will be able to simultaneously fit the frequency and energy dependence of the lag.  Furthermore, we have also only explored the simplest lamppost type geometry for the irradiating X-ray source.  This is almost certainly far too simplistic, and models using an extended X-ray corona (both radially and vertically) will provide a more realistic approach.  

\section*{Acknowledgements}
ACF thanks the Royal Society for support. DRW is supported by a CITA National Fellowship

\bibliographystyle{mn2e}
\bibliography{agn}

\begin{thebibliography}{}

\bibitem[\protect\citeauthoryear{{Alston}, {Vaughan} \& {Uttley}}{{Alston}
  et~al.}{2013}]{alston13}
{Alston} W.~N.,  {Vaughan} S.,    {Uttley} P.,  2013, \mnras, 435, 1511

\bibitem[\protect\citeauthoryear{{Bardeen}, {Press} \& {Teukolsky}}{{Bardeen}
  et~al.}{1972}]{bardeen72}
{Bardeen} J.~M.,  {Press} W.~H.,    {Teukolsky} S.~A.,  1972, \apj, 178, 347

\bibitem[\protect\citeauthoryear{{Bentz}, {Denney}, {Cackett} et~al.,}{{Bentz}
  et~al.}{2006}]{bentz06}
{Bentz} M.~C.,  {Denney} K.~D.,  {Cackett} E.~M.,    et~al., 2006, \apj, 651,
  775

\bibitem[\protect\citeauthoryear{{Bentz} et~al.,}{{Bentz}
  et~al.}{2009}]{bentz09}
{Bentz} M.~C.,  et~al., 2009, \apj, 705, 199

\bibitem[\protect\citeauthoryear{{Bentz}, {Horne}, {Barth}, {Bennert},
  {Canalizo}, {Filippenko}, {Gates}, {Malkan}, {Minezaki}, {Treu}, {Woo} \&
  {Walsh}}{{Bentz} et~al.}{2010}]{bentz10}
{Bentz} M.~C.,  {Horne} K.,  {Barth} A.~J.,  {Bennert} V.~N.,  {Canalizo} G.,
  {Filippenko} A.~V.,  {Gates} E.~L.,  {Malkan} M.~A.,  {Minezaki} T.,  {Treu}
  T.,  {Woo} J.-H.,    {Walsh} J.~L.,  2010, \apjl, 720, L46

\bibitem[\protect\citeauthoryear{{Bentz}, {Peterson}, {Netzer}, {Pogge} \&
  {Vestergaard}}{{Bentz} et~al.}{2009}]{bentz09_rl}
{Bentz} M.~C.,  {Peterson} B.~M.,  {Netzer} H.,  {Pogge} R.~W.,
  {Vestergaard} M.,  2009, \apj, 697, 160

\bibitem[\protect\citeauthoryear{{Blandford} \& {McKee}}{{Blandford} \&
  {McKee}}{1982}]{blandmckee82}
{Blandford} R.~D.,  {McKee} C.~F.,  1982, \apj, 255, 419

\bibitem[\protect\citeauthoryear{{Brenneman} \& {Reynolds}}{{Brenneman} \&
  {Reynolds}}{2006}]{brenneman06}
{Brenneman} L.~W.,  {Reynolds} C.~S.,  2006, \apj, 652, 1028

\bibitem[\protect\citeauthoryear{{Cackett}, {Fabian}, {Zogbhi}, {Kara},
  {Reynolds} \& {Uttley}}{{Cackett} et~al.}{2013}]{cackett13}
{Cackett} E.~M.,  {Fabian} A.~C.,  {Zogbhi} A.,  {Kara} E.,  {Reynolds} C.,
  {Uttley} P.,  2013, \apjl, 764, L9

\bibitem[\protect\citeauthoryear{{Campana} \& {Stella}}{{Campana} \&
  {Stella}}{1995}]{campana95}
{Campana} S.,  {Stella} L.,  1995, \mnras, 272, 585

\bibitem[\protect\citeauthoryear{{Cassatella}, {Uttley}, {Wilms} \&
  {Poutanen}}{{Cassatella} et~al.}{2012}]{cassatella11}
{Cassatella} P.,  {Uttley} P.,  {Wilms} J.,    {Poutanen} J.,  2012, \mnras,
  422, 2407

\bibitem[\protect\citeauthoryear{{Crummy}, {Fabian}, {Gallo} \&
  {Ross}}{{Crummy} et~al.}{2006}]{crummy06}
{Crummy} J.,  {Fabian} A.~C.,  {Gallo} L.,    {Ross} R.~R.,  2006, \mnras, 365,
  1067

\bibitem[\protect\citeauthoryear{{Das}, {Crenshaw}, {Hutchings}, {Deo},
  {Kraemer}, {Gull}, {Kaiser}, {Nelson} \& {Weistrop}}{{Das}
  et~al.}{2005}]{das05}
{Das} V.,  {Crenshaw} D.~M.,  {Hutchings} J.~B.,  {Deo} R.~P.,  {Kraemer}
  S.~B.,  {Gull} T.~R.,  {Kaiser} M.~E.,  {Nelson} C.~H.,    {Weistrop} D.,
  2005, \aj, 130, 945

\bibitem[\protect\citeauthoryear{{De Marco}, {Ponti}, {Cappi}, {Dadina},
  {Uttley}, {Cackett}, {Fabian} \& {Miniutti}}{{De Marco}
  et~al.}{2013}]{demarco13}
{De Marco} B.,  {Ponti} G.,  {Cappi} M.,  {Dadina} M.,  {Uttley} P.,  {Cackett}
  E.~M.,  {Fabian} A.~C.,    {Miniutti} G.,  2013, \mnras, 431, 2441

\bibitem[\protect\citeauthoryear{{De Marco}, {Ponti}, {Uttley}, {Cappi},
  {Dadina}, {Fabian} \& {Miniutti}}{{De Marco} et~al.}{2011}]{demarco11}
{De Marco} B.,  {Ponti} G.,  {Uttley} P.,  {Cappi} M.,  {Dadina} M.,  {Fabian}
  A.~C.,    {Miniutti} G.,  2011, \mnras, 417, L98

\bibitem[\protect\citeauthoryear{{Emmanoulopoulos}, {McHardy} \&
  {Papadakis}}{{Emmanoulopoulos} et~al.}{2011}]{emma11}
{Emmanoulopoulos} D.,  {McHardy} I.~M.,    {Papadakis} I.~E.,  2011, \mnras,
  416, L94

\bibitem[\protect\citeauthoryear{{Fabian}, {Kara} et~al.,}{{Fabian}
  et~al.}{2013}]{fabian13}
{Fabian} A.~C.,  {Kara} E.,    et~al., 2013, \mnras, 429, 2917

\bibitem[\protect\citeauthoryear{{Fabian}, {Rees}, {Stella} \&
  {White}}{{Fabian} et~al.}{1989}]{fabian89}
{Fabian} A.~C.,  {Rees} M.~J.,  {Stella} L.,    {White} N.~E.,  1989, \mnras,
  238, 729

\bibitem[\protect\citeauthoryear{{Fabian}, {Zoghbi}, {Ross}, {Uttley}, {Gallo},
  {Brandt}, {Blustin}, {Boller}, {Caballero-Garcia}, {Larsson}, {Miller},
  {Miniutti}, {Ponti}, {Reis}, {Reynolds}, {Tanaka} \& {Young}}{{Fabian}
  et~al.}{2009}]{fabian09}
{Fabian} A.~C.,  {Zoghbi} A.,  {Ross} R.~R.,  {Uttley} P.,  {Gallo} L.~C.,
  {Brandt} W.~N.,  {Blustin} A.~J.,  {Boller} T.,  {Caballero-Garcia} M.~D.,
  {Larsson} J.,  {Miller} J.~M.,  {Miniutti} G.,  {Ponti} G.,  {Reis} R.~C.,
  {Reynolds} C.~S.,  {Tanaka} Y.,    {Young} A.~J.,  2009, \nat, 459, 540

\bibitem[\protect\citeauthoryear{{George} \& {Fabian}}{{George} \&
  {Fabian}}{1991}]{georgefabian91}
{George} I.~M.,  {Fabian} A.~C.,  1991, \mnras, 249, 352

\bibitem[\protect\citeauthoryear{{Gierli{\'n}ski} \& {Done}}{{Gierli{\'n}ski}
  \& {Done}}{2004}]{gierlinski04}
{Gierli{\'n}ski} M.,  {Done} C.,  2004, \mnras, 349, L7

\bibitem[\protect\citeauthoryear{{Kara}, {Fabian}, {Cackett}, {Miniutti} \&
  {Uttley}}{{Kara} et~al.}{2013b}]{kara13b}
{Kara} E.,  {Fabian} A.~C.,  {Cackett} E.~M.,  {Miniutti} G.,    {Uttley} P.,
  2013b, \mnras, 430, 1408

\bibitem[\protect\citeauthoryear{{Kara}, {Fabian}, {Cackett}, {Steiner},
  {Uttley}, {Wilkins} \& {Zoghbi}}{{Kara} et~al.}{2013a}]{kara13a}
{Kara} E.,  {Fabian} A.~C.,  {Cackett} E.~M.,  {Steiner} J.~F.,  {Uttley} P.,
  {Wilkins} D.~R.,    {Zoghbi} A.,  2013a, \mnras, 428, 2795

\bibitem[\protect\citeauthoryear{{Kara}, {Fabian}, {Cackett}, {Uttley},
  {Wilkins} \& {Zoghbi}}{{Kara} et~al.}{2013c}]{kara13c}
{Kara} E.,  {Fabian} A.~C.,  {Cackett} E.~M.,  {Uttley} P.,  {Wilkins} D.~R.,
   {Zoghbi} A.,  2013c, \mnras, 434, 1129

\bibitem[\protect\citeauthoryear{{Kaspi}, {Smith}, {Netzer}, {Maoz}, {Jannuzi}
  \& {Giveon}}{{Kaspi} et~al.}{2000}]{kaspi00}
{Kaspi} S.,  {Smith} P.~S.,  {Netzer} H.,  {Maoz} D.,  {Jannuzi} B.~T.,
  {Giveon} U.,  2000, \apj, 533, 631

\bibitem[\protect\citeauthoryear{{Kotov}, {Churazov} \& {Gilfanov}}{{Kotov}
  et~al.}{2001}]{kotov01}
{Kotov} O.,  {Churazov} E.,    {Gilfanov} M.,  2001, \mnras, 327, 799

\bibitem[\protect\citeauthoryear{{Laor}}{{Laor}}{1991}]{laor91}
{Laor} A.,  1991, \apj, 376, 90

\bibitem[\protect\citeauthoryear{{Legg}, {Miller}, {Turner}, {Giustini},
  {Reeves} \& {Kraemer}}{{Legg} et~al.}{2012}]{legg12}
{Legg} E.,  {Miller} L.,  {Turner} T.~J.,  {Giustini} M.,  {Reeves} J.~N.,
  {Kraemer} S.~B.,  2012, \apj, 760, 73

\bibitem[\protect\citeauthoryear{{Miller}}{{Miller}}{2007}]{miller07}
{Miller} J.~M.,  2007, \araa, 45, 441

\bibitem[\protect\citeauthoryear{{Miller}, {Reynolds}, {Fabian}, {Miniutti} \&
  {Gallo}}{{Miller} et~al.}{2009}]{miller09}
{Miller} J.~M.,  {Reynolds} C.~S.,  {Fabian} A.~C.,  {Miniutti} G.,    {Gallo}
  L.~C.,  2009, \apj, 697, 900

\bibitem[\protect\citeauthoryear{{Miller}, {Turner}, {Reeves} \&
  {Braito}}{{Miller} et~al.}{2010}]{lancemiller10}
{Miller} L.,  {Turner} T.~J.,  {Reeves} J.~N.,    {Braito} V.,  2010, \mnras,
  408, 1928

\bibitem[\protect\citeauthoryear{{Miyamoto} \& {Kitamoto}}{{Miyamoto} \&
  {Kitamoto}}{1989}]{miyamoto89}
{Miyamoto} S.,  {Kitamoto} S.,  1989, \nat, 342, 773

\bibitem[\protect\citeauthoryear{{Nandra}, {George}, {Mushotzky}, {Turner} \&
  {Yaqoob}}{{Nandra} et~al.}{1997}]{nandra97}
{Nandra} K.,  {George} I.~M.,  {Mushotzky} R.~F.,  {Turner} T.~J.,    {Yaqoob}
  T.,  1997, \apjl, 488, L91

\bibitem[\protect\citeauthoryear{{Nandra}, {O'Neill}, {George} \&
  {Reeves}}{{Nandra} et~al.}{2007}]{nandra07}
{Nandra} K.,  {O'Neill} P.~M.,  {George} I.~M.,    {Reeves} J.~N.,  2007,
  \mnras, 382, 194

\bibitem[\protect\citeauthoryear{{Nayakshin} \& {Kazanas}}{{Nayakshin} \&
  {Kazanas}}{2002}]{nayakshin02}
{Nayakshin} S.,  {Kazanas} D.,  2002, \apj, 567, 85

\bibitem[\protect\citeauthoryear{{Nowak}, {Vaughan}, {Wilms}, {Dove} \&
  {Begelman}}{{Nowak} et~al.}{1999}]{nowak99}
{Nowak} M.~A.,  {Vaughan} B.~A.,  {Wilms} J.,  {Dove} J.~B.,    {Begelman}
  M.~C.,  1999, \apj, 510, 874

\bibitem[\protect\citeauthoryear{{Papadakis}, {Nandra} \&
  {Kazanas}}{{Papadakis} et~al.}{2001}]{papadakis01}
{Papadakis} I.~E.,  {Nandra} K.,    {Kazanas} D.,  2001, \apjl, 554, L133

\bibitem[\protect\citeauthoryear{{Peterson} et~al.,}{{Peterson}
  et~al.}{2004}]{petersonetal04}
{Peterson} B.~M.,  et~al., 2004, \apj, 613, 682

\bibitem[\protect\citeauthoryear{{Poutanen}}{{Poutanen}}{2002}]{poutanen02}
{Poutanen} J.,  2002, \mnras, 332, 257

\bibitem[\protect\citeauthoryear{{Reynolds}}{{Reynolds}}{1997}]{reynolds97}
{Reynolds} C.~S.,  1997, \mnras, 286, 513

\bibitem[\protect\citeauthoryear{{Reynolds} \& {Nowak}}{{Reynolds} \&
  {Nowak}}{2003}]{reynoldsnowak03}
{Reynolds} C.~S.,  {Nowak} M.~A.,  2003, \physrep, 377, 389

\bibitem[\protect\citeauthoryear{{Reynolds}, {Young}, {Begelman} \&
  {Fabian}}{{Reynolds} et~al.}{1999}]{reynolds99}
{Reynolds} C.~S.,  {Young} A.~J.,  {Begelman} M.~C.,    {Fabian} A.~C.,  1999,
  \apj, 514, 164

\bibitem[\protect\citeauthoryear{{Ross} \& {Fabian}}{{Ross} \&
  {Fabian}}{2005}]{rossfabian05}
{Ross} R.~R.,  {Fabian} A.~C.,  2005, \mnras, 358, 211

\bibitem[\protect\citeauthoryear{Shapiro}{Shapiro}{1964}]{shapiro64}
{Shapiro} I.~I.,  1964, Phys. Rev. Letters, 13, 789

\bibitem[\protect\citeauthoryear{{Tanaka}, {Nandra}, {Fabian}, {Inoue},
  {Otani}, {Dotani}, {Hayashida}, {Iwasawa}, {Kii}, {Kunieda}, {Makino} \&
  {Matsuoka}}{{Tanaka} et~al.}{1995}]{tanaka95}
{Tanaka} Y.,  {Nandra} K.,  {Fabian} A.~C.,  {Inoue} H.,  {Otani} C.,  {Dotani}
  T.,  {Hayashida} K.,  {Iwasawa} K.,  {Kii} T.,  {Kunieda} H.,  {Makino} F.,
   {Matsuoka} M.,  1995, \nat, 375, 659

\bibitem[\protect\citeauthoryear{{Thorne}}{{Thorne}}{1974}]{thorne74}
{Thorne} K.~S.,  1974, \apj, 191, 507

\bibitem[\protect\citeauthoryear{{Tripathi}, {Misra}, {Dewangan} \&
  {Rastogi}}{{Tripathi} et~al.}{2011}]{tripathi11}
{Tripathi} S.,  {Misra} R.,  {Dewangan} G.,    {Rastogi} S.,  2011, \apjl, 736,
  L37

\bibitem[\protect\citeauthoryear{{Uttley}, {Wilkinson}, {Cassatella}, {Wilms},
  {Pottschmidt}, {Hanke} \& {B{\"o}ck}}{{Uttley} et~al.}{2011}]{uttley11}
{Uttley} P.,  {Wilkinson} T.,  {Cassatella} P.,  {Wilms} J.,  {Pottschmidt} K.,
   {Hanke} M.,    {B{\"o}ck} M.,  2011, \mnras, 414, L60

\bibitem[\protect\citeauthoryear{{Vestergaard}}{{Vestergaard}}{2002}]{vestergaard02}
{Vestergaard} M.,  2002, \apj, 571, 733

\bibitem[\protect\citeauthoryear{{Wilkins} \& {Fabian}}{{Wilkins} \&
  {Fabian}}{2013}]{wilkins13}
{Wilkins} D.~R.,  {Fabian} A.~C.,  2013, \mnras, 430, 247

\bibitem[\protect\citeauthoryear{{Zoghbi} \& {Fabian}}{{Zoghbi} \&
  {Fabian}}{2011}]{zog11_rej1034}
{Zoghbi} A.,  {Fabian} A.~C.,  2011, \mnras, 418, 2642

\bibitem[\protect\citeauthoryear{{Zoghbi}, {Fabian}, {Reynolds} \&
  {Cackett}}{{Zoghbi} et~al.}{2012}]{zoghbi12}
{Zoghbi} A.,  {Fabian} A.~C.,  {Reynolds} C.~S.,    {Cackett} E.~M.,  2012,
  \mnras, 422, 129

\bibitem[\protect\citeauthoryear{{Zoghbi}, {Fabian}, {Uttley}, {Miniutti},
  {Gallo}, {Reynolds}, {Miller} \& {Ponti}}{{Zoghbi} et~al.}{2010}]{zoghbi10}
{Zoghbi} A.,  {Fabian} A.~C.,  {Uttley} P.,  {Miniutti} G.,  {Gallo} L.~C.,
  {Reynolds} C.~S.,  {Miller} J.~M.,    {Ponti} G.,  2010, \mnras, 401, 2419

\bibitem[\protect\citeauthoryear{{Zoghbi}, {Reynolds}, {Cackett}, {Miniutti},
  {Kara} \& {Fabian}}{{Zoghbi} et~al.}{2013}]{zoghbi13}
{Zoghbi} A.,  {Reynolds} C.,  {Cackett} E.~M.,  {Miniutti} G.,  {Kara} E.,
  {Fabian} A.~C.,  2013, \apj, 767, 121

\bibitem[\protect\citeauthoryear{{Zoghbi}, {Uttley} \& {Fabian}}{{Zoghbi}
  et~al.}{2011}]{zog11_1h0707}
{Zoghbi} A.,  {Uttley} P.,    {Fabian} A.~C.,  2011, \mnras, 412, 59

\end{thebibliography}

\end{document}